\documentclass[final,5p,times,twocolumn]{elsarticle}
\makeatletter
\def\ps@pprintTitle{%
  \let\@oddhead\@empty
  \let\@evenhead\@empty
  \def\@oddfoot{}%
  \let\@evenfoot\@oddfoot}
\makeatother

\usepackage[utf8]{inputenc}
\usepackage[T1]{fontenc}

\usepackage{graphicx}
\usepackage{booktabs}
\usepackage{array}
\usepackage{amsmath,amssymb}
\usepackage[expansion=false]{microtype}
\usepackage{url}
\usepackage{xcolor}
\usepackage{algorithm}
\usepackage{algpseudocode}
\usepackage[labelfont=bf,labelsep=period]{caption}
\usepackage[colorlinks=true,
            linkcolor=blue!60!black,
            urlcolor=blue!60!black,
            citecolor=blue!60!black,
            pdfborder={0 0 0}]{hyperref}

\biboptions{numbers,sort&compress}

\newcommand{\chg}[1]{#1}
\newcommand{\chgold}[1]{#1}

\begin{document}

\begin{frontmatter}

\title{RespCCT and FreqCCT: Compact, Explainable and Cost-Aware Convolutional Transformers for Multi-Class Respiratory Disease Classification}

\author[a]{Auntor Chakma\corref{cor1}}
\ead{auntorchakma@gmail.com}

\author[b]{Mst. Farhana Akter}
\ead{farhanaaktermst@gmail.com}

\author[c]{Abu Henaf Rashid Ahmad Mufti}
\ead{muftiahmad963@gmail.com}

\author[b]{Asmana Yasmin}
\ead{asmanayasmin@gmail.com}

\affiliation[a]{organization={ELITE Research Lab},
                city={New York},
                state={NY},
                country={USA}}

\affiliation[b]{organization={Department of Computer Science and Engineering, East Delta University},
                addressline={Abdullah Al Noman Road, Noman Society, East Nasirabad, Khulshi},
                city={Chattogram},
                postcode={4209},
                country={Bangladesh}}

\affiliation[c]{organization={Department of Computer Science and Engineering, United International University},
                addressline={United City, Madani Avenue, Badda},
                city={Dhaka},
                postcode={1212},
                country={Bangladesh}}

\cortext[cor1]{Corresponding author.}

\begin{abstract}
\chg{Chest auscultation depends on the listener, and software that classifies lung sounds could give a second reading where trained listeners are scarce. The most accurate published classifiers fine-tune large pretrained networks, and their accuracy is rarely reported with running cost or paired statistics. This study makes three contributions. (i) It proposes RespCCT and FreqCCT, two compact convolutional transformers of 173,187 and 124,999 parameters that learn from random initialization on mel, constant-Q and cepstral images of 5 s frames from three public databases, ICBHI, KAUH and RD@TR, with 1,665 audio files in seven classes. (ii) It compares both models with five ImageNet-pretrained networks and larger distilled variants on one test set, using paired bootstrap intervals. (iii) It measures the cost of every pipeline stage on one CPU thread, tests whether attribution maps depend on the trained weights, and probes whether the learned features reveal the recording source. On the test fold, RespCCT reaches an accuracy of 98.13\% and a macro F1 of 0.970, and FreqCCT reaches 97.58\% and 0.951. Over five folds, RespCCT with three seeds reaches 97.65\% and 0.959, and FreqCCT with one seed reaches 96.95\% and 0.941. The two strongest pretrained networks reach a macro F1 of 0.984, whereas RespCCT needs 2.0 to 64.6 times less weight storage. Computing the three input images takes longer than running either model, and a linear probe identifies the source database with 0.958 accuracy. A team can weigh accuracy against storage and time before device testing.}
\end{abstract}


\end{frontmatter}

\section{Introduction}
\label{sec:intro}

\chg{Chest auscultation is usually the first examination of a patient with a breathing complaint, and its value depends on who is listening. This matters because chronic respiratory diseases were the third leading cause of death worldwide in 2019, with 4.0 million deaths and 454.6 million prevalent cases, and chronic obstructive pulmonary disease (COPD) accounted for 3.3 million of those deaths \cite{GBD2023}. In one study, 185 physicians and medical students matched 24 recorded respiratory sounds from children to standard descriptions, and only the pulmonologists scored significantly better than the other groups \cite{Hafke2019}. A study at four teaching hospitals found the same gradient, with an overall accuracy of 60.3\% for medical students, 53.4\% for interns, 68.8\% for residents and 80.1\% for fellows \cite{Kim2021}. Software that classifies a recorded sound could provide a second reading at the point of care, for example on a phone paired with a digital stethoscope or on a small computer in a community clinic \cite{Huang2023}.}

\chg{Published classifiers of lung sounds largely share one design, in which the waveform is converted into an image of its energy over time and frequency and a convolutional network assigns the image to a class. Huang et al.\ describe this combination as the most common one in the field \cite{Huang2023}. The strongest results on the ICBHI database come from large networks, most of them pretrained on images or audio. On the official split, where each breathing cycle is labeled as normal, crackle, wheeze or both, the ICBHI Score rose from 58.29\% for a co-tuned ResNet-50 with ImageNet weights \cite{Nguyen2022} to 62.37\% for an Audio Spectrogram Transformer pretrained on ImageNet and AudioSet \cite{Bae2023}, and to 66.55\% for a multi-view spectrogram transformer \cite{He2024MVST}. These models were built for accuracy, whereas a screening tool also has to run on inexpensive hardware, and Huang et al.\ note that the price of electronic stethoscopes already limits their use in low-resource areas \cite{Huang2023}. Compact models for diagnostic labels exist. Roy et al.\ pooled ICBHI, KAUH and RD@TR for seven classes, reported 98.88\% accuracy and ran their network on a Raspberry Pi 4 \cite{Roy2024}, and Park et al.\ ran a 264 K-parameter CNN on a Raspberry Pi 5 with a neural processing unit \cite{Park2025}. Size is only one requirement, because Rudin argues that high-stakes decisions call for models whose reasoning can be checked \cite{Rudin2019}.}

\chg{This progress leaves five questions open for a team that has to choose and deploy a model. First, it is unclear how close a compact transformer can come to pretrained networks when it learns from lung sounds alone. Compact convolutional transformers (CCTs) reach 98\% accuracy on CIFAR-10 from scratch with 3.7 M parameters \cite{Hassani2021}, but, to our knowledge, they have not been tested on pooled respiratory databases, and the compact respiratory models above were not compared with pretrained networks through paired intervals. Second, accuracy is normally reported without its cost and without paired statistics, so a difference of one or two percent cannot be told from noise (Section~\ref{sec:litdb}). Third, it has not been tested, to our knowledge, whether knowledge distillation, in which a small student model is trained to match the outputs of teacher models \cite{Hinton2015,Gou2021}, can pass pretrained knowledge to a compact transformer trained on lung sounds. Fourth, attribution maps are usually shown without a check that they depend on the trained model, and the recording source can act as a shortcut that no map reveals \cite{Geirhos2020}, which matters here because three of the seven classes come from a single database. Fifth, models are called small because their parameter count is low, while the conversion of sound into images runs on the same device and is left out of the cost, although on a microcontroller the MFCC computation was estimated at 763 ms against 127 ms for the network that reads it \cite{Abadade2024}.}

\chg{The present study answers the five questions within one experimental design. A total of 1,665 audio files from ICBHI, KAUH and RD@TR were pooled into seven diagnostic classes and cut into 5 s frames, and each frame was represented by a mel spectrogram, a constant-Q transform and cepstral coefficients, the same three views that Roy et al.\ used \cite{Roy2024}. Two CCTs, RespCCT and FreqCCT, were trained on these frames from random initialization, because a convolutional tokenizer allows a transformer to learn from small datasets without pretraining \cite{Hassani2021}. Five ImageNet-pretrained networks, two larger CCT variants and their distilled versions were then scored on the same test frames, and every difference between two models was given a paired bootstrap interval. The same trained models were examined with a randomization test, occlusion and Shapley attributions and a probe for the source database, and each stage of the pipeline was timed on one CPU thread. Because the pooled data tie some classes to one source, the accuracy is reported together with the source probe, which measures how much of it could come from the source.}

\chg{Addressing these questions, this paper makes the following contributions.}
\begin{itemize}
\item \chg{\textbf{Two compact models.} RespCCT with 173,187 parameters and 0.69 MB and FreqCCT with 124,999 parameters and 0.50 MB learn from random initialization on three stacked time--frequency images of each 5 s frame. On seven pooled classes RespCCT reaches an accuracy of 98.13\% and a macro F1 of 0.970 on the test fold, and 97.65\% and 0.959 over five folds and three seeds.}
\item \chg{\textbf{A paired comparison.} Five ImageNet-pretrained networks, two larger CCT variants and an ensemble are scored on the same 2,359 test frames, and every difference carries a file-level bootstrap interval. MobileViT XXS and EfficientNet-B0 lead RespCCT with accuracies of 99.11\% and 99.15\%, and ShuffleNetV2, MobileNetV3 Small and ResNet-18 cannot be separated from it.}
\item \chg{\textbf{A test of knowledge distillation.} Two pretrained teachers, MobileViT XXS and MobileNetV3 Small, were distilled into CCT-M with 967,619 parameters, which reaches an accuracy of 98.64\% and a macro F1 of 0.975. Its accuracy gain over RespCCT is 0.0051 with a 95\% interval of $-$0.0004 to $+$0.011, so it is not reliable, and the distilled runs also change the mixup and label smoothing.}
\item \chg{\textbf{Checks on the explanations.} The class-weighted rollout loses its rank correlation with the original map, from 0.95 to 0.17, when the trained weights are randomized, occlusion and Shapley maps differ between classes with mean profile correlations of 0.29 and 0.28, and a linear probe names the source database from the 64-dimensional embedding with an accuracy of 0.958, against 0.582 for always naming the largest source.}
\item \chg{\textbf{A cost account of the whole pipeline.} Parameters, operations, weight storage, activation memory, per-transform timing, and PyTorch and ONNX Runtime latency are reported together. The three input images take 29.6 ms per frame, of which the constant-Q transform takes 25.0 ms, against 6.43 ms for the RespCCT network.}
\end{itemize}
\chg{Neither model is the most accurate in the comparison. MobileViT XXS and EfficientNet-B0 reach accuracies of 99.11\% and 99.15\% and a macro F1 of 0.984, against 98.13\% and 0.970 for RespCCT, and the intervals of these gaps in macro F1 exclude zero. RespCCT and FreqCCT need no pretrained weights, they store 0.69 MB and 0.50 MB, and RespCCT has 5.5 times fewer parameters than MobileViT XXS. On one CPU thread the RespCCT network takes 6.43 ms per 5 s frame without its view gate and the whole pipeline takes 36.0 ms, of which 29.6 ms is spent on the three input images. Their cost is reported for every stage, so that a team with limited hardware can choose a model size with the trade-offs in view.}

\begin{figure*}[!t]
\centering
\includegraphics[width=\textwidth]{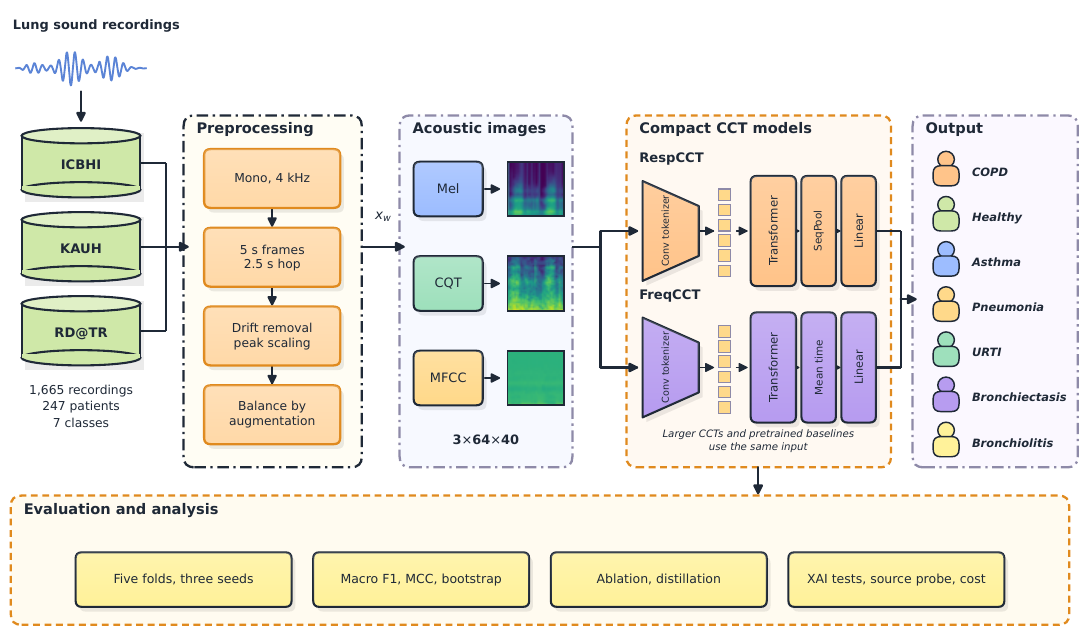}
\caption{\chgold{Overview of the study, covering data pooling and framing, three time--frequency images per frame, the compact and reference models and the evaluation steps.}}
\label{fig:frame}
\end{figure*}

\chg{The remainder of the paper follows the five questions. Section~\ref{sec:lit} groups earlier work by the problem it leaves open and compares the studies reviewed in Table~\ref{tab:lit}. Section~\ref{sec:method} describes the three databases with their class and source mix, the preprocessing and the three input images, the two proposed models, the pretrained baselines, the training and distillation recipe, and the statistical, explanation and cost tests. Section~\ref{sec:results} reports the results in the order of the questions, Section~\ref{sec:disc} discusses the choice of device, the explanation checks and the choice between a compact and a pretrained model, Section~\ref{sec:limits} gives the limitations and future work, and Section~\ref{sec:concl} concludes. Figure~\ref{fig:frame} gives an overview of the study.}

\section{Related work}
\label{sec:lit}

\subsection{\chg{Databases, labels and metrics}}
\label{sec:litdb}
\chg{Results from different studies can be compared only when it is clear what each database labels and how each score was obtained, and the three collections used here differ on both counts. ICBHI holds 920 annotated recordings, about 5.5 hours, from 126 subjects, and its 6,898 respiratory cycles are labeled for crackles and wheezes, with a diagnosis for each patient \cite{Rocha2019}. KAUH contains chest-wall recordings of 112 subjects, 35 healthy and 77 with conditions such as asthma, heart failure and COPD, each exported with three stethoscope filters \cite{Fraiwan2021}, and RD@TR provides 12-channel recordings labeled by five COPD severities \cite{Altan2020}. Labels at the level of a breathing cycle and at the level of a patient define two different tasks, because a wheeze can occur in asthma, in COPD and in other conditions. A model that finds wheezes therefore does not name the disease, and results for the two tasks cannot be compared. Metrics and splits vary as well. On the official ICBHI split, cycle-level studies report the ICBHI Score, the mean of sensitivity and specificity, and published values rose from 58.29\% for a co-tuned ResNet-50 in 2022 \cite{Nguyen2022} to 66.55\% for a multi-view spectrogram transformer in 2024 \cite{He2024MVST}. Diagnostic studies more often report accuracy on random splits, with 99.01\% for a dual-channel CNN--LSTM trained on MFCC features of 5,054 augmented ICBHI samples \cite{Zhang2024}, and 99.23\% on ICBHI and 98.48\% on KAUH with RD@TR for binary COPD detection \cite{Dong2026}. Data quality adds variation, since removing poor recordings raised diagnostic accuracy from 88\% to 97\% in one interstitial lung disease cohort but not in a second \cite{Fava2024}. None of these studies reports paired intervals between models on a shared test set.}

\subsection{\chg{Pretrained and compact transformers}}
\chg{Whether a transformer can learn from a few thousand lung sound frames depends on how far it relies on pretraining. Vision transformers trained on mid-sized datasets score a few percent below ResNets of comparable size, and they overtake them only after pretraining on 14 M to 300 M images \cite{Dosovitskiy2021}. The Audio Spectrogram Transformer \cite{Gong2021} and its respiratory adaptations start from ImageNet and AudioSet weights \cite{Bae2023,Kim2024}. Compact convolutional transformers reduce this dependence with a convolutional tokenizer and sequence pooling, reaching 98\% accuracy on CIFAR-10 from scratch with 3.7 M parameters, and variants as small as 0.28 M parameters \cite{Hassani2021}. These results come from images, and the present study tests the CCT design on pooled lung sounds.}

\subsection{\chg{Compact respiratory models and their cost}}
\chg{Interest in small respiratory models is older than the present transformer designs, yet the comparisons that justify them against pretrained networks rarely carry paired statistics or a count of what the input images cost. Acharya and Basu combined a CNN with a recurrent network, reached an ICBHI Score of 66.31\% on four classes, and cut memory fourfold by weight quantization without loss of performance \cite{Acharya2020}. Roy et al.\ pooled three databases for seven classes and reported 98.88\% accuracy in five-fold cross-validation with 0.414 M parameters, above that of fine-tuned ResNet-50, MobileNet and ShuffleNetV2, and their network needed 3.48 s per segment on a Raspberry Pi 4 \cite{Roy2024}. EasyNet uses two convolution layers on 3 s frames and reports accuracies of 0.997 on ICBHI, 1.0 on KAUH and 0.998 on both combined \cite{Hassan2024}. Quantization \cite{Jacob2018} and microcontroller-scale design \cite{Lin2020} shrink such models further, but image cost is seldom separated from network cost. On an Arduino Nano 33 BLE Sense, the MFCC computation of a normal-versus-asthma classifier was estimated at 763 ms, against 127 ms for the CNN that reads it \cite{Abadade2024}. The present study adds paired intervals and times each input image separately.}

\subsection{\chg{Multiple input representations}}
\chg{Studies that treat the time--frequency image as a design variable have compared single representations, patches of one spectrogram and stacks of several views, and each choice implies a different cost. Pham et al.\ compared log-mel, gammatone, stacked MFCC and constant-Q spectrograms one at a time and found constant-Q weakest \cite{Pham2021}. MVST fuses patches of several shapes from one mel spectrogram with a gate \cite{He2024MVST}, while Park et al.\ stacked mel, chroma and MFCC images for a 264 K-parameter CNN that reached 89.46\% accuracy on a Hailo-8 accelerator attached to a Raspberry Pi 5 \cite{Park2025}. Metadata also help, since text on sex, age, device and body location raised the best ICBHI Score by 1.17\% \cite{Kim2024BTS}, and sex and age improved contrastive pretraining \cite{Moummad2023}. Roy et al.\ fused mel, constant-Q and MFCC views in separate streams and found that networks reading one or two of the views were less accurate than the three-view network \cite{Roy2024}. To our knowledge, none of them reports the computation time of each transform next to the accuracy lost when it is removed, which Section~\ref{sec:abl} measures.}

\subsection{\chg{Knowledge distillation}}
\chg{A small model can inherit pretrained knowledge without carrying the pretrained network, which is the aim of knowledge distillation and the reason it suits devices with limited memory. Hinton et al.\ compressed the knowledge of an ensemble into a single model that is easier to deploy \cite{Hinton2015}, and a survey groups later variants by the knowledge transferred, the training scheme and the teacher--student architecture \cite{Gou2021}. For respiratory sounds, Pham et al.\ distilled a 4.5 M-parameter teacher into a 0.6 M-parameter student that matched its ICBHI Score of 0.84 on three disease classes \cite{Pham2021}, and soft labels from an ensemble raised the ICBHI Score of a single model to 64.39\% \cite{Toikkanen2025}. Both studies used ICBHI alone, whereas here two small pretrained networks teach compact transformers trained from scratch on pooled databases (Section~\ref{sec:training}).}

\subsection{\chg{Explanation checks and shortcut cues}}
\chg{Attribution maps are easy to produce and easy to over-read, and a map that looks plausible to a clinician may not depend on what the model learned. Adebayo et al.\ found some saliency methods independent of both the model and the data, so their maps barely change when trained weights are randomized \cite{Adebayo2018}, and reviews disagree on how far such maps should be trusted in high-stakes use \cite{Samek2021,Rudin2019}. A second risk lies in the data. Shortcuts are decision rules that score well on standard benchmarks but fail to transfer \cite{Geirhos2020}, and for lung sounds the stethoscope and the recording site are such cues. RespireNet fine-tuned an ImageNet-pretrained ResNet-34 separately for each recording device and reached an ICBHI Score of 56.20\% \cite{Gairola2021}, and stethoscope-guided contrastive training with gradient reversal reached 61.71\% \cite{Kim2024}. Both reduce the device effect in training, but neither measures the source information left in the learned features, which the present study measures alongside a randomization check on the maps.}

\subsection{\chg{Summary of open issues}}
\chg{Taken together, the studies above leave five issues open, and each one matches a question of Section~\ref{sec:intro}. Compact transformers have not been tested on pooled respiratory databases, and compact respiratory models are rarely scored against pretrained networks on one test set. Published results come from different protocols and are reported without paired intervals, so small differences cannot be judged. Distillation into compact transformers trained from random initialization on pooled databases has not been reported. Attribution maps are seldom checked against randomized weights, and the source information in the learned features is not measured. Finally, the cost of the input images is rarely counted next to that of the network, although it can exceed it. Table~\ref{tab:lit} compares the studies reviewed in this section.}

\begin{table*}[!t]
\centering\footnotesize
\setlength{\tabcolsep}{3pt}
\caption{\chg{Thirteen studies on respiratory sound classification. Results are quoted as reported and are not comparable across class sets, metrics and splits.}}
\label{tab:lit}
\begin{tabular}{>{\raggedright\arraybackslash}p{0.6cm}>{\raggedright\arraybackslash}p{1.8cm}>{\raggedright\arraybackslash}p{3.1cm}>{\raggedright\arraybackslash}p{3.5cm}>{\raggedright\arraybackslash}p{3.1cm}>{\raggedright\arraybackslash}p{2.9cm}}
\toprule
\chg{Year} & \chgold{Study} & \chgold{Method} & \chgold{Reported result} & \chgold{Limitation} & \chgold{Addressed here}\\
\midrule
\chg{2020} & \chgold{Acharya and Basu \cite{Acharya2020}} & \chgold{CNN--RNN, patient-specific tuning, weight quantization} & \chgold{ICBHI Score 66.31\%, memory cut fourfold} & \chgold{No comparison with pretrained networks} & \chgold{Same data and metrics for all models}\\
\chg{2021} & \chgold{Gairola et al.\ \cite{Gairola2021}} & \chgold{RespireNet, ResNet34 with ImageNet weights, device-specific fine-tuning} & \chgold{ICBHI Score 56.20\%} & \chgold{Needs ImageNet weights} & \chgold{Training from random initialization}\\
\chg{2022} & \chgold{Nguyen and Pernkopf \cite{Nguyen2022}} & \chgold{Fine-tuned ResNets, co-tuning, device correction} & \chgold{ICBHI Score 58.29\%} & \chgold{Needs ImageNet weights, source information not measured} & \chgold{Source probe on learned features}\\
\chg{2023} & \chgold{Bae et al.\ \cite{Bae2023}} & \chgold{Pretrained AST, patch-mix contrastive learning} & \chgold{ICBHI Score 62.37\%} & \chgold{Large pretrained model} & \chgold{Training from random initialization}\\
\chg{2024} & \chgold{He et al.\ \cite{He2024MVST}} & \chgold{Multi-view patches of one mel spectrogram, gated fusion} & \chgold{ICBHI Score 66.55\%} & \chgold{One transform, cost not reported} & \chgold{Three transforms, removed one at a time}\\
\chg{2024} & \chgold{Zhang et al.\ \cite{Zhang2024}} & \chgold{Dual-channel CNN--LSTM on MFCC} & \chgold{Accuracy 99.01\% (ICBHI, random split)} & \chgold{One input representation} & \chgold{Paired comparison on one test set}\\
\chg{2024} & \chg{Roy et al. \cite{Roy2024}} & \chg{triple-scale self-operational network, seven pooled classes} & \chg{accuracy 98.88\% (five-fold), 0.414 M parameters, 0.0055 GFLOPs, 3.48 s per segment on a Raspberry Pi 4} & \chg{compared with fine-tuned ResNet-50, MobileNet and ShuffleNetV2 without paired intervals, transform time not separated} & \chg{paired intervals and per-transform timing}\\
\chg{2024} & \chgold{Kim et al.\ \cite{Kim2024}} & \chgold{Stethoscope-guided supervised contrastive learning with gradient reversal} & \chgold{ICBHI Score 61.71\%} & \chgold{Device effect reduced in training, not measured in the learned features} & \chgold{Source probe on learned features}\\
\chg{2024} & \chg{Abadade et al. \cite{Abadade2024}} & \chg{TinyML pipeline on an Arduino Nano 33 BLE Sense} & \chg{MFCC estimated at 763 ms against 127 ms for the CNN} & \chg{estimated timing, two classes} & \chg{measured timing of all three images and every model}\\
\chg{2024} & \chg{Hassan et al. \cite{Hassan2024}} & \chg{two-layer CNN on 3 s frames} & \chg{accuracy 0.997 (ICBHI), 1.0 (KAUH), 0.998 (combined)} & \chg{no comparison with pretrained networks} & \chg{same test set and metrics for all models}\\
\chg{2024} & \chg{Kim et al. \cite{Kim2024BTS}} & \chg{text--audio model with metadata} & \chg{ICBHI Score 1.17\% above the previous best} & \chg{large pretrained model} & \chg{training from random initialization}\\
\chg{2025} & \chg{Park et al. \cite{Park2025}} & \chg{mel, chroma and MFCC stacked, 264 K-parameter CNN} & \chg{accuracy 89.46\%, Raspberry Pi 5 with Hailo-8} & \chg{compared with five baselines, ResNet50 among them, without paired intervals} & \chg{paired bootstrap intervals on one test set}\\
\chg{2026} & \chgold{Dong et al.\ \cite{Dong2026}} & \chgold{CNN, BiLSTM and vision transformer hybrid, binary COPD detection} & \chgold{Accuracy 99.23\% (ICBHI), 98.48\% (KAUH with RD@TR)} & \chgold{Two classes only} & \chgold{Seven pooled classes}\\
\bottomrule
\end{tabular}
\end{table*}

\section{Methodology}
\label{sec:method}

\subsection{Dataset description and class distribution}
\label{sec:block1}
\chgold{The data were pooled from three public databases so that the models meet more than one recording device and cohort, namely ICBHI \cite{Rocha2019}, KAUH \cite{Fraiwan2021} and RD@TR \cite{Altan2020}.} \chgold{Seven diagnostic classes were kept, namely COPD, healthy, asthma, pneumonia, upper respiratory tract infection (URTI), bronchiectasis and bronchiolitis.} \chg{Recordings with any other diagnosis were not used. ICBHI also holds lower respiratory tract infection, but only two recordings, and a class of two recordings cannot be divided between training, validation and test folds \cite{Zhang2024}.} In all, 1,665 \chgold{audio files} from 247 patients gave 11,902 frames of 5 s (Table~\ref{tab:data}). \chgold{The pool is strongly unbalanced, since COPD supplies 9,837 frames (83\%), mostly from ICBHI and RD@TR. Class and source are also tied.} \chgold{URTI, bronchiectasis and bronchiolitis come only from ICBHI, 513 of the 520 asthma frames come from KAUH, and RD@TR contributes only COPD. A rule that assigns each source its most frequent class would reach 86.3\% accuracy (10,272 of 11,902 frames), and Section~\ref{sec:probe} tests how much of this source information the models absorb.} \chgold{The three rarest classes together have 52 \chgold{audio files} from 27 patients.}

\begin{table*}[!t]
\centering\small
\caption{\chgold{Composition of the pooled data. Frames are 5 s windows before augmentation, and test frames are those of fold 0.}}
\label{tab:data}
\begin{tabular}{lrrrrl}
\toprule
Class & \chgold{Audio files} & Patients & Frames & Test frames & Frames by source\\
\midrule
COPD & 1,324 & 115 & 9,837 & 1,967 & ICBHI 6,055, RD@TR 3,629, KAUH 153\\
Healthy & 140 & 61 & 828 & 139 & KAUH 588, ICBHI 240\\
Asthma & 97 & 33 & 520 & 101 & KAUH 513, ICBHI 7\\
Pneumonia & 52 & 11 & 355 & 70 & ICBHI 259, KAUH 96\\
URTI & 23 & 14 & 159 & 35 & ICBHI 159\\
Bronchiectasis & 16 & 7 & 112 & 29 & ICBHI 112\\
Bronchiolitis & 13 & 6 & 91 & 18 & ICBHI 91\\
\midrule
Total & 1,665 & 247 & 11,902 & 2,359 & \\
\bottomrule
\end{tabular}
\end{table*}

\subsection{Audio preprocessing and data augmentation}
\chg{Audio files differ in device, length and loudness, so every audio file passed through the same cleaning, which is close to the preprocessing that Roy et al.\ reported for the same databases \cite{Roy2024}.} \chgold{\chgold{Each audio file} was converted to mono and resampled to 4 kHz, which covers the 0 to 2 kHz band used here.} \chgold{Audio files} shorter than 5 s were zero padded. \chgold{Frames of 5 s (20,000 samples) were cut every 2.5 s, which gives 50\% overlap.} \chg{The mean breathing cycle in ICBHI lasts 2.7 s \cite{Nguyen2022}, so a 5 s frame holds on average 1.9 cycles, enough to contain a full inspiration and expiration while keeping the input image small. The frame length was fixed beforehand and was not compared with other lengths in this study.} \chgold{In each frame the DFT bins from 0 to 1 Hz were set to zero to remove offset and slow drift, and the frame was divided by its largest absolute sample, so that loudness could not decide the class (Table~\ref{tab:params}).}

\chgold{Balancing was done on the pooled frames before the folds were drawn. Augmented copies of every class other than COPD were generated until each class matched COPD, which gave 9,837 frames per class and 68,859 in all. The five folds were then drawn from this pool, stratified by class, and only the original frames of the test fold were kept for testing (2,359 frames for fold 0). For fold 0, the four training folds held 55,087 frames, of which 5,509, 10\%, were drawn at random for validation with the same class proportions and including augmented copies, and 49,578 were used for training.} Each copy applied one of four equally likely operations. These were time stretching with a rate from 0.85 to 1.15, pitch shifting by up to $\pm$3 semitones, additive Gaussian noise with amplitude factor 0.001 to 0.01, or all three in sequence. \chgold{Test frames were never augmented.} The only augmentation during training was the time masking of Section~\ref{sec:training}.

\subsection{Feature extraction with mel, CQT and MFCC}
\begin{figure*}[!t]
\centering
\includegraphics[width=\textwidth]{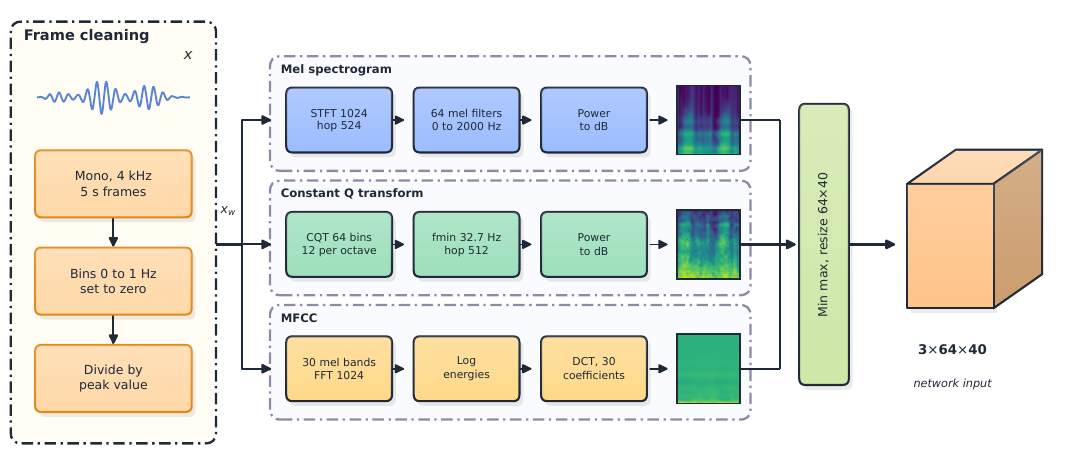}
\caption{\chgold{Preprocessing and image generation. A cleaned 5 s frame is converted into mel, CQT and MFCC images, which are scaled, resized and stacked.}}
\label{fig:s2i}
\end{figure*}

\begin{figure}[!t]
\centering
\includegraphics[width=\columnwidth]{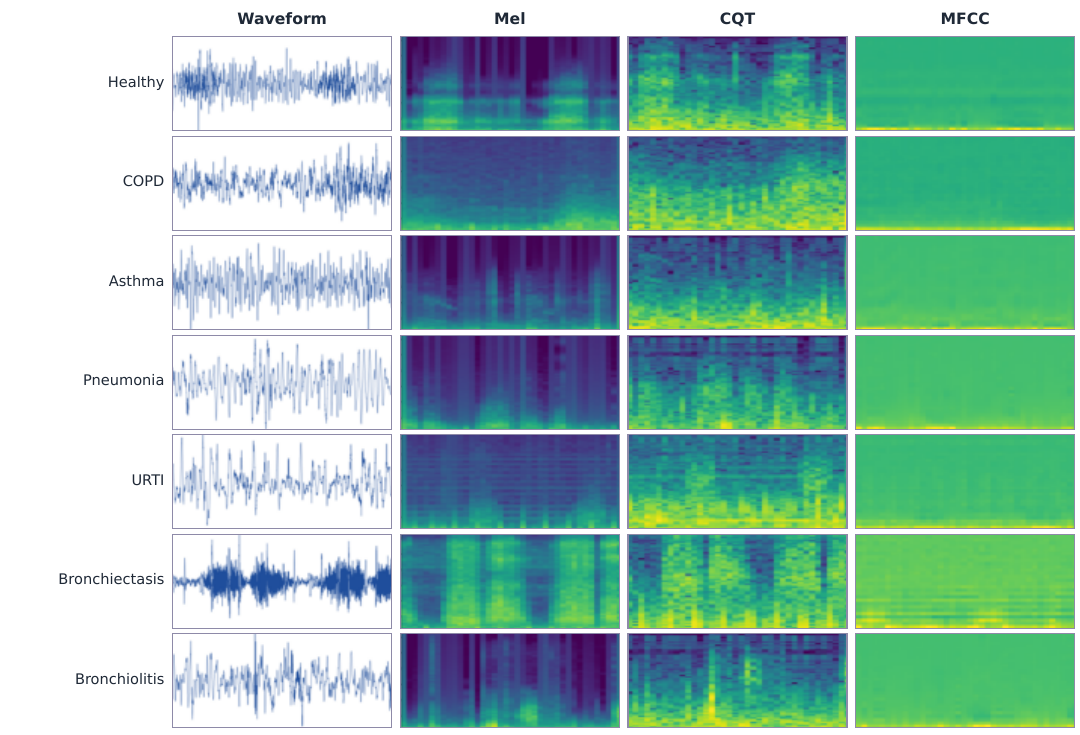}
\caption{\chgold{One real frame per class, with its waveform and the three images before resizing.}}
\label{fig:ex}
\end{figure}

\chgold{Each frame is turned into three images of how its energy is distributed over time and frequency, and the stack of three is the network input, much as color planes form a photograph.} \chg{The three views describe the same frame in different ways. The mel spectrogram gives a perceptually spaced view of the whole 0 to 2 kHz band, the constant-Q transform gives frequency bins of equal ratio that are dense at low frequencies, where the mel filters are coarse, and the cepstral coefficients describe the spectral envelope in 30 numbers per column. The band of 0 to 2 kHz contains the respiratory sound range of about 60 to 2,000 Hz \cite{Molaie2015}. Table~\ref{tab:params} gives the reason for every setting, and Section~\ref{sec:abl} tests whether all three views are needed.} Figure~\ref{fig:s2i} gives the stages, Figure~\ref{fig:ex} shows one real frame per class, and Table~\ref{tab:params} lists the parameters. \chgold{Features were computed with librosa \cite{McFee2015} (the version preinstalled in Google Colab at the time of the runs), with its default centring, which pads each frame by $N/2$ samples at both ends.}

\chgold{The mel spectrogram starts from the short-time Fourier transform.} \chgold{For a padded frame $\tilde x[n]$ and a Hamming window $w[n]$ of length $N=1024$ with hop $H$ (Eq.~\eqref{eq:stft}),}
\begin{equation}
\label{eq:stft}X[k,t]=\sum_{n=0}^{N-1}w[n]\,\tilde x[n+tH]\,e^{-j2\pi kn/N},\qquad k=0,\dots,N/2,
\end{equation}
\chgold{with $t=0,\dots,T-1$ and $T=1+\lfloor L/H\rfloor$, which gives 39 columns for $H=524$ and 40 for $H=512$. The power spectrum is pooled by $M=64$ triangular filters $H_m[k]$ between 0 and 2 kHz on the Slaney mel scale (linear below 1 kHz and logarithmic above), expressed in decibels relative to the maximum and clipped 80 dB below it (Eq.~\eqref{eq:mel}),}
\begin{equation}
\label{eq:mel}\begin{split}P_m[t]&=\sum_k H_m[k]\,|X[k,t]|^2,\\ S_{\mathrm{mel}}[m,t]&=\max\!\Big(10\log_{10}\frac{P_m[t]}{\max_{m',t'}P_{m'}[t']},\,-80\Big).\end{split}
\end{equation}
\chgold{The constant-Q transform (CQT) uses windows whose length shrinks as frequency rises, so that every bin has the same ratio $Q$ of center frequency to bandwidth (Eq.~\eqref{eq:cqt}),}
\begin{equation}
\label{eq:cqt}\begin{split}X_{\mathrm{cq}}[k,t]&=\frac{1}{\sqrt{N_k}}\sum_{n=0}^{N_k-1}w_{N_k}[n]\,\tilde x[n+tH_q]\,e^{-j2\pi Qn/N_k},\\ f_k&=f_{\min}2^{k/B},\quad Q=\frac{1}{2^{1/B}-1},\quad N_k=\Big\lceil\frac{Qf_s}{f_k}\Big\rceil,\end{split}
\end{equation}
\chgold{with $k=0,\dots,63$, $f_{\min}=32.7$ Hz, $B=12$ bins per octave and $H_q=512$. The 64 bins reach about 1.24 kHz and are dense at low frequencies, where the mel filters are coarse.} \chgold{The image is $10\log_{10}|X_{\mathrm{cq}}|^2$ relative to its maximum and clipped at $-80$ dB, as in Eq.~\eqref{eq:mel}.} \chgold{The mel frequency cepstral coefficients (MFCC) \cite{Davis1980} apply an orthonormal discrete cosine transform to the log energies of a 30-band mel filter bank with the same window, hop 512 and 0 to 2 kHz range (Eq.~\eqref{eq:mfcc}),}
\begin{equation}
\label{eq:mfcc}c[n,t]=\gamma_n\sum_{m=1}^{30}\ell_m[t]\cos\!\Big(\frac{\pi n}{30}\big(m-\tfrac12\big)\Big),\qquad n=0,\dots,29,
\end{equation}
\chgold{where $\ell_m[t]$ is the band energy in decibels clipped 80 dB below its maximum, $\gamma_0=\sqrt{1/30}$ and $\gamma_n=\sqrt{2/30}$ for $n\geq1$. All 30 coefficients are kept.} \chgold{Each image $S$ is scaled to $[0,1]$ by $\tilde S=(S-\min S)/(\max S-\min S+10^{-8})$ and resized by bilinear interpolation to 64$\times$40, and the three images are stored in half precision.}   \chgold{Algorithm~\ref{alg:pipeline} summarizes the complete path from \chgold{an audio file} to a class.}

\begin{algorithm}[!t]
\caption{Inference path from an audio file to class probabilities}
\label{alg:pipeline}
\small
\begin{algorithmic}[1]
\Require audio file $r$, trained model $f_\theta$, either RespCCT or FreqCCT
\Ensure class probabilities $\hat p_j\in[0,1]^7$ for every frame $j$ of $r$
\State convert $r$ to mono, resample to $f_s=4$ kHz, zero pad to at least 5 s \Comment{input}
\State cut frames $x_j$ of $L=20{,}000$ samples with a hop of 10,000 samples \Comment{framing}
\For{each frame $x_j$}
  \State set the DFT bins from 0 to 1 Hz to zero and invert the DFT \Comment{cleaning}
  \State $x_j\gets x_j/\max_n|x_j[n]|$
  \State $S_{\mathrm{mel}}\gets$ Eq.~\eqref{eq:mel} \Comment{images}
  \State $S_{\mathrm{cq}}\gets$ Eq.~\eqref{eq:cqt}
  \State $S_{\mathrm{mfcc}}\gets$ Eq.~\eqref{eq:mfcc}
  \For{$S\in\{S_{\mathrm{mel}},S_{\mathrm{cq}},S_{\mathrm{mfcc}}\}$}
    \State $\tilde S\gets(S-\min S)/(\max S-\min S+10^{-8})$, resized to $64\times40$
  \EndFor
  \State $X_j\gets[\tilde S_{\mathrm{mel}};\tilde S_{\mathrm{cq}};\tilde S_{\mathrm{mfcc}}]\in[0,1]^{3\times64\times40}$
  \State $\hat p_j\gets f_\theta(X_j)$ \Comment{Eqs.~\eqref{eq:gate}--\eqref{eq:pool}}
\EndFor
\State \Return $\{\hat p_j\}$, and for a file-level label the most frequent $\arg\max_k\hat p_{j,k}$
\end{algorithmic}
\end{algorithm}

\begin{table*}[!t]
\centering\small
\caption{\chg{Preprocessing, augmentation and image parameters, with the reason for each setting.}}
\label{tab:params}
\begin{tabular}{p{1.7cm}p{3.0cm}p{4.1cm}p{5.7cm}}
\toprule
Stage & Item & Value & \chg{Reason}\\
\midrule
Signal & Sampling rate, channels & 4 kHz, mono & \chg{keeps the 0 to 2 kHz band analyzed here and gives 20,000 samples per 5 s frame}\\
 & Frame length, hop & 5 s (20,000 samples), 2.5 s (10,000 samples) & \chg{a 5 s frame holds on average 1.9 breathing cycles of ICBHI with a mean cycle of 2.7 s \cite{Nguyen2022}, and 50\% overlap doubles the number of frames without new recordings}\\
 & Drift removal & DFT bins 0 to 1 Hz set to zero & \chg{removes offset and slow drift below the band of respiratory sounds}\\
 & Scaling & Division by the maximum absolute value & \chg{removes loudness differences between devices, so that loudness cannot decide the class}\\
Augmentation & Stretch, pitch, noise & 0.85 to 1.15, $\pm$3 semitones, factor 0.001 to 0.01 & \chg{each copy changes tempo, pitch or background noise within these ranges and keeps the label of the frame it was made from}\\
Mel view & Bands, FFT, hop, window & 64 (Slaney), 1024 (256 ms), 524 (131 ms), Hamming, 0 to 2000 Hz & \chg{64 rows match the 64 CQT bins, 1024 points give a resolution of 3.9 Hz, the hop gives 39 columns close to the 40 of the other views, and the band is the full range available at 4 kHz}\\
 & \chg{Level} & \chg{dB with an 80 dB floor} & \chg{limits the weight of near-silent bins}\\
CQT view & Bins, per octave, lowest frequency, hop & 64, 12, 32.7 Hz, 512 (128 ms) & \chg{one bin per semitone up to about 1.24 kHz, dense at low frequencies where the mel filters are coarse, and 40 columns per frame}\\
MFCC view & Coefficients, mel bands, FFT, hop & 30, 30, 1024, 512, orthonormal DCT-II & \chg{every coefficient of a 30-band bank is kept, so no cepstral detail is discarded, and the hop gives 40 columns}\\
Image & Scaling, size, storage & Min--max to [0, 1] per view, bilinear to 64$\times$40, 3$\times$64$\times$40, float16 & \chg{the same value range for every view, one grid for stacking the three views, and half the storage of 32-bit floats}\\
\bottomrule
\end{tabular}
\end{table*}

\subsection{Proposed model architectures}

\subsubsection{RespCCT}
\begin{figure*}[!t]
\centering
\includegraphics[width=\textwidth]{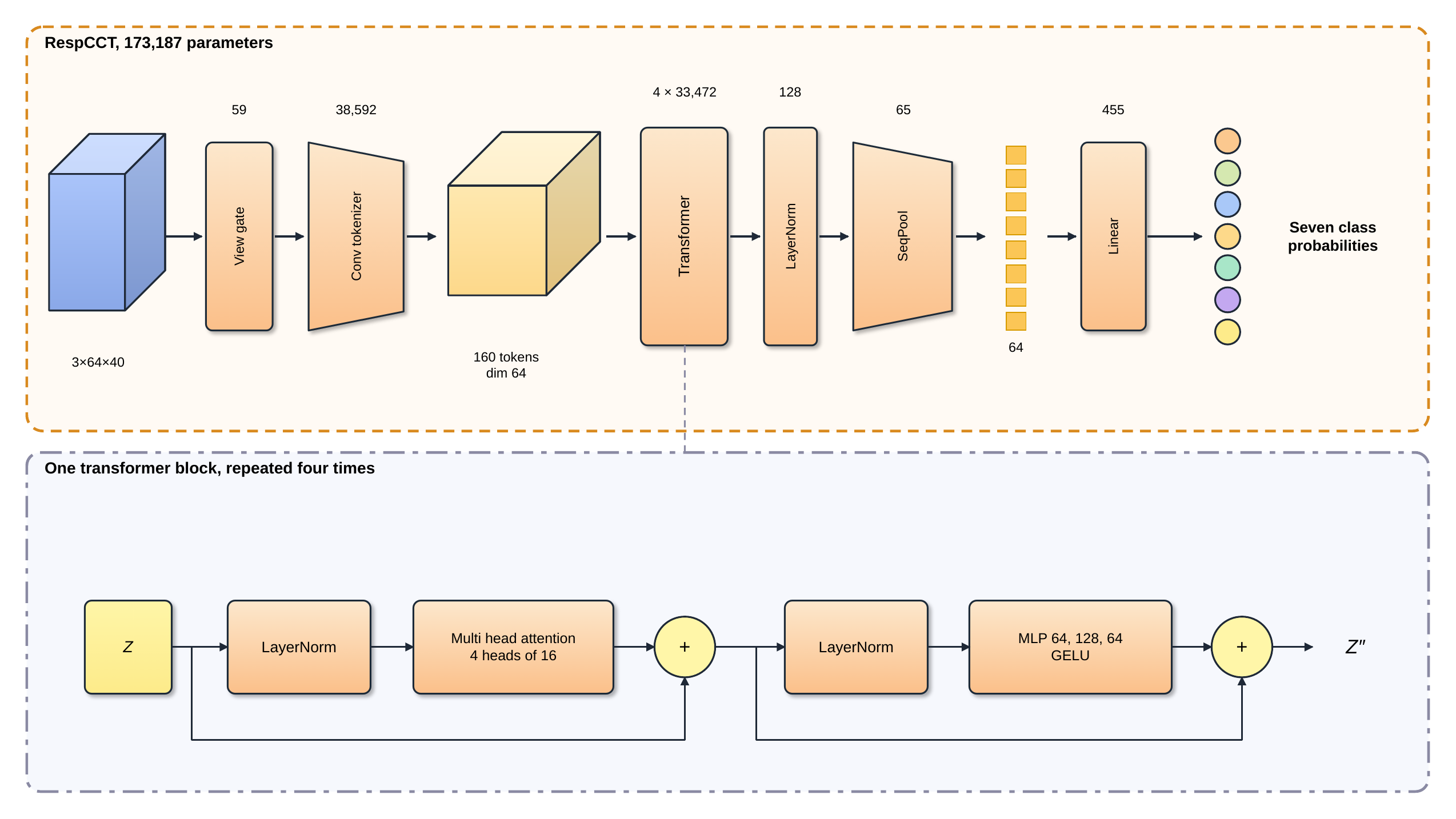}
\caption{\chgold{RespCCT architecture with the parameter count of each stage, and the lower panel expands one transformer block.}}
\label{fig:arch}
\end{figure*}

\chgold{RespCCT is a compact convolutional transformer \cite{Hassani2021} (Figure~\ref{fig:arch}, Table~\ref{tab:arch}).} \chgold{It replaces the patch cutting of a vision transformer \cite{Dosovitskiy2021} with a small convolutional tokenizer, which suits small datasets, and it was kept small on purpose so that it can run where large pretrained networks cannot.}

\chgold{\emph{View gate.} The three views are first reweighted, because they are not equally informative.} \chgold{With $\bar{x}\in\mathbb{R}^3$ the spatial mean of each view, $\sigma$ the logistic sigmoid and $\rho$ the ReLU (Eq.~\eqref{eq:gate}),}
\begin{equation}
\label{eq:gate}g=2\,\sigma\big(W_2\,\rho(W_1\bar{x}+b_1)+b_2\big)\in(0,2)^3,\qquad \hat X_c=g_cX_c,
\end{equation}
where $W_1\in\mathbb{R}^{8\times3}$ and $W_2\in\mathbb{R}^{3\times8}$. The factor 2 lets the gate raise a view as well as lower it. The gate has 59 parameters.

\chgold{\emph{Convolutional tokenizer.} Two stages of 3$\times$3 convolution (64 channels, with bias), ReLU and 3$\times$3 max pooling with stride 2 reduce the 64$\times$40 plane to 16$\times$10.} \chgold{The map is read row by row into 160 tokens $z_i\in\mathbb{R}^{64}$. Each token already sees its neighbors through the convolutions, so no positional embedding is added.}

\chgold{\emph{Transformer encoder.} Four pre-norm blocks follow.} \chgold{For tokens $Z\in\mathbb{R}^{160\times64}$, a block computes (Eq.~\eqref{eq:block})}
\begin{equation}
\label{eq:block}\begin{split}Z'&=Z+\mathrm{MHA}(\mathrm{LN}(Z)),\\ Z''&=Z'+W_4\,\mathrm{GELU}(W_3\,\mathrm{LN}(Z')+b_3)+b_4,\end{split}
\end{equation}
\chgold{with multi-head attention \cite{Vaswani2017} $\mathrm{MHA}(Y)=[h_1;\dots;h_4]W^O$ and $h_i=\mathrm{softmax}\big(YW_i^Q(YW_i^K)^\top/\sqrt{16}\big)YW_i^V$, that is, four heads of width 16 and a feed-forward width of 128.} Stochastic depth randomly skips blocks during training.

\chgold{\emph{Sequence pooling and classifier.} A linear layer scores each output token of the fourth block, and a softmax over the 160 scores gives pooling weights.} \chgold{The frame representation and class probabilities are (Eq.~\eqref{eq:pool})}
\begin{equation}
\label{eq:pool}a_i=\frac{\exp(u^\top z_i+b_u)}{\sum_{j=1}^{160}\exp(u^\top z_j+b_u)},\quad r=\sum_{i=1}^{160}a_iz_i,\quad \hat p=\mathrm{softmax}(W_cr+b_c),
\end{equation}
\chgold{with $W_c\in\mathbb{R}^{7\times64}$. RespCCT has 173,187 parameters (173,128 without the gate) and needs 124.2 MFLOPs per frame.}

\begin{table}[!t]
\centering\small
\caption{\chgold{RespCCT layers and parameters (input 3$\times$64$\times$40).}}
\label{tab:arch}
\begin{tabular}{p{2.2cm}p{3.9cm}r}
\toprule
\textbf{Stage} & \textbf{Operation and output} & \textbf{Parameters}\\
\midrule
\chgold{View gate} & \chgold{Linear(3,8), ReLU, Linear(8,3), $2\sigma$} & \chgold{59}\\
\chgold{Tokenizer} & \chgold{2$\times$[Conv 3$\times$3 (64), ReLU, MaxPool 3/2/1], 160$\times$64} & \chgold{38,720}\\
Encoder & 4 blocks of LN, MHA (4 heads), LN, MLP 64, 128, 64 & 4$\times$33,472\\
Sequence pooling & Linear(64,1), softmax over tokens & 65\\
Classifier & Linear(64,7) & 455\\
\midrule
\textbf{Total} & & \textbf{173,187}\\
\bottomrule
\end{tabular}
\end{table}

\subsubsection{FreqCCT}
\begin{figure*}[!t]
\centering
\includegraphics[width=\textwidth]{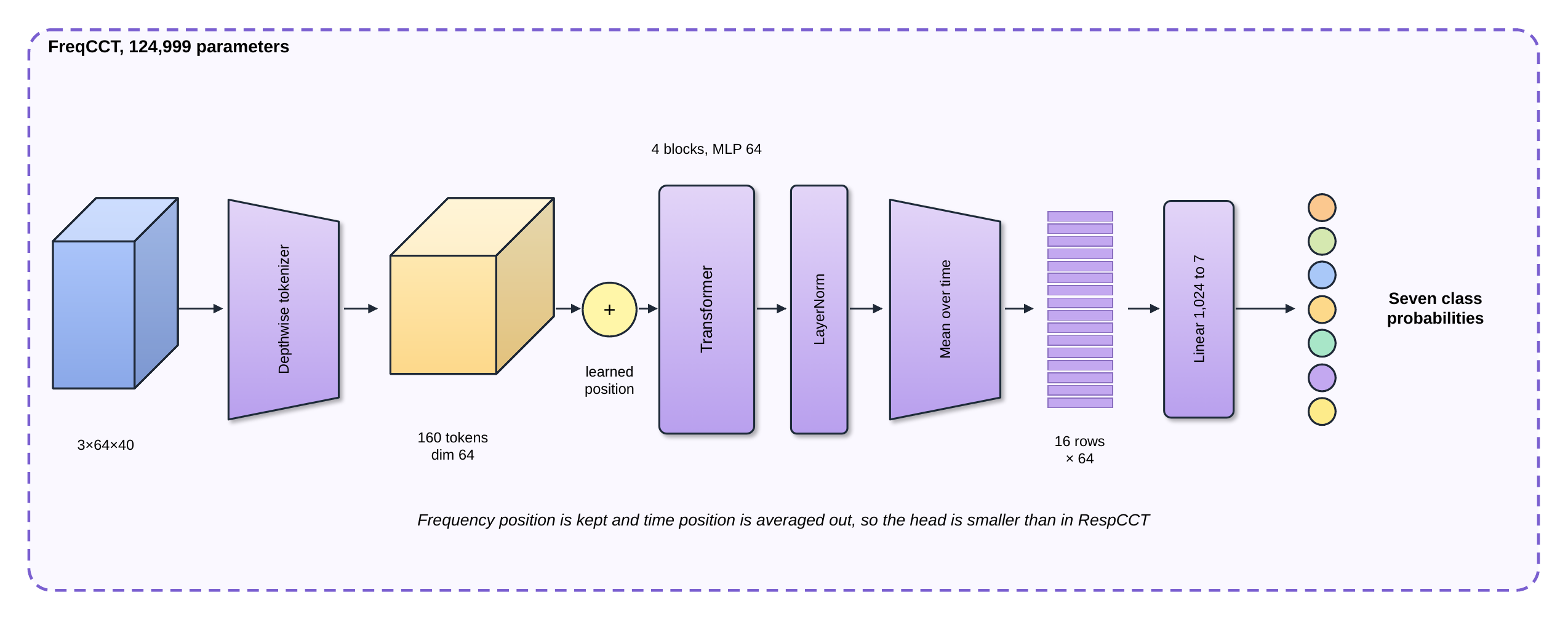}
\caption{\chgold{FreqCCT architecture, in which frequency rows are kept and time is averaged before the classifier.}}
\label{fig:freq}
\end{figure*}

\chgold{FreqCCT is the second model, designed for the smallest budgets.} \chgold{It keeps the tokenizer and encoder of RespCCT with four changes (Figure~\ref{fig:freq}).} \chgold{The second convolution of the tokenizer is depthwise separable, a learned positional embedding is added to the 160 tokens, and the feed-forward width is 64 instead of 128.} \chgold{The head also differs. After a final layer norm the tokens are averaged over the 10 time positions, the 16 frequency rows are kept, and a linear layer maps the resulting $16\times64=1{,}024$ values to the seven classes.} \chgold{Frequency position was kept and time position averaged out because RespCCT changed few predictions under circular shifts in time (Section~\ref{sec:expl}).} \chgold{FreqCCT has no view gate, 124,999 parameters and 72.5 MFLOPs per frame, 42\% fewer operations than RespCCT.}

\subsubsection{Larger CCT variants}
\chgold{Two larger variants test whether more capacity, with or without pretrained teachers, closes the gap to the pretrained networks.} \chgold{CCT-M (embedding 128, six blocks, four heads) has 967,619 parameters and 536.8 MFLOPs, and CCT-L (embedding 192, six blocks, six heads) has 2,151,811 parameters and 1,135.5 MFLOPs.} Both use a learned positional embedding. \chgold{An ensemble averages the outputs of CCT-M and CCT-L, each with and without distillation, for 6.2 M parameters and 3,345 MFLOPs.} \chgold{Table~\ref{tab:sizes} lists the size of every model.}

\begin{table*}[!t]
\centering\small
\caption{\chgold{Size of every model, given as parameters, MFLOPs per 5 s frame and, for the CCT family, embedding width, number of blocks and number of heads. Distilled students have the size of their plain versions, and the teachers are MobileViT XXS and MobileNetV3 Small.}}
\label{tab:sizes}
\begin{tabular}{llcrrc}
\toprule
Model & Group & Initialization & Parameters & MFLOPs & Width / blocks / heads\\
\midrule
\chgold{RespCCT} & \chgold{Proposed} & \chgold{Random} & \chgold{173,187} & \chgold{124.2} & \chgold{64 / 4 / 4}\\
\chgold{FreqCCT} & \chgold{Proposed} & \chgold{Random} & \chgold{124,999} & \chgold{72.5} & \chgold{64 / 4 / 4}\\
\chgold{CCT-M} & \chgold{Larger variant} & \chgold{Random} & \chgold{967,619} & \chgold{536.8} & \chgold{128 / 6 / 4}\\
\chgold{CCT-M + KD} & \chgold{Distilled student} & \chgold{Random} & \chgold{967,619} & \chgold{536.8} & \chgold{128 / 6 / 4}\\
\chgold{CCT-L} & \chgold{Larger variant} & \chgold{Random} & \chgold{2,151,811} & \chgold{1,135.5} & \chgold{192 / 6 / 6}\\
\chgold{CCT-L + KD} & \chgold{Distilled student} & \chgold{Random} & \chgold{2,151,811} & \chgold{1,135.5} & \chgold{192 / 6 / 6}\\
\chgold{Ensemble of four CCTs} & \chgold{Ensemble} & \chgold{Random} & \chgold{6.2 M} & \chgold{3,345} & \chgold{n/a}\\
\chgold{ShuffleNetV2 $\times$0.5} & \chgold{Pretrained baseline} & \chgold{ImageNet} & \chgold{348,967} & \chgold{17.2} & \chgold{n/a}\\
\chgold{MobileNetV3 Small} & \chgold{Pretrained baseline, teacher} & \chgold{ImageNet} & \chgold{931,047} & \chgold{24.5} & \chgold{n/a}\\
\chgold{MobileViT XXS} & \chgold{Pretrained baseline, teacher} & \chgold{ImageNet} & \chgold{953,271} & \chgold{109.6} & \chgold{n/a}\\
\chgold{EfficientNet-B0} & \chgold{Pretrained baseline} & \chgold{ImageNet} & \chgold{4,016,515} & \chgold{168.2} & \chgold{n/a}\\
\chgold{ResNet-18} & \chgold{Pretrained baseline} & \chgold{ImageNet} & \chgold{11,180,103} & \chgold{773.8} & \chgold{n/a}\\
\bottomrule
\end{tabular}
\end{table*}

\subsection{Pretrained baseline models}
\chgold{Five networks were initialized from ImageNet weights and fine-tuned end to end on the same frames, as the reference against which the compact models are judged, namely ShuffleNetV2 $\times$0.5 \cite{Ma2018} (348,967 parameters), MobileNetV3 Small \cite{Howard2019} (931,047), MobileViT XXS \cite{Mehta2022} (953,271), EfficientNet-B0 \cite{Tan2019} (4,016,515) and ResNet-18 \cite{He2016} (11,180,103).} \chgold{ShuffleNetV2 and MobileNetV3 Small are the smallest widely used ImageNet networks, MobileViT XXS combines convolutions with vision transformer blocks \chg{(the larger MobileViT reaches 78.4\% top-1 accuracy on ImageNet with about 6 M parameters \cite{Mehta2022})}, and EfficientNet-B0 and ResNet-18 are common larger references.} The input was normalized with ImageNet statistics and enlarged to 128$\times$80 by bilinear interpolation. \chgold{Data, balancing, label smoothing, schedule, time masking, weight averaging, batch size and stopping rule matched those of RespCCT. For the two smallest networks the learning rate was chosen on the validation split from $\{10^{-4},3\cdot10^{-4},10^{-3}\}$, and for the others it was set to $10^{-3}$.}

\subsection{Training configuration and knowledge distillation}
\label{sec:training}
\chgold{RespCCT and FreqCCT share one recipe, fixed by a search on the validation split of the training folds.} \chgold{It uses AdamW \cite{Loshchilov2019} with learning rate 0.002946, weight decay 0, batch size 128, label smoothing $\varepsilon=0.15$, a cosine schedule and gradient-norm clipping at 1.0, with stochastic depth 0.0847, MLP dropout 0.0877 and attention dropout 0.0811.} \chgold{Each training frame receives two time masks \cite{Park2019}, each zeroing up to 8 of the 40 columns in all three views.} \chgold{The weights are averaged after every step as $\bar\theta\gets\delta_s\bar\theta+(1-\delta_s)\theta$ with $\delta_s=\min\big(0.99,(1+s)/(10+s)\big)$, and the averaged weights are the ones validated and reported.} \chgold{Training stops after 100 epochs or after 12 epochs in which the validation loss does not fall by more than $10^{-4}$, and the weights at the last such improvement are kept.} \chgold{The objective is cross-entropy with smoothed labels $\tilde y_k=(1-\varepsilon)\mathbb{1}[k=y]+\varepsilon/7$ and no class weights, since the training pool is balanced.} 

Distillation tests whether pretrained knowledge can reach a model that has none. \chgold{Two teachers, MobileViT XXS and MobileNetV3 Small, were trained on the same folds, and their softmax outputs are averaged into $\bar p_T$. With $\hat p$ the student output (Eq.~\eqref{eq:kd}),}
\begin{equation}
\label{eq:kd}\mathcal{L}=(1-\alpha)\,\mathcal{L}_{\mathrm{CE}}+\alpha\,\mathrm{KL}(\bar p_T\,\|\,\hat p),\qquad \alpha=0.8,
\end{equation}
\chgold{at temperature $T=1$ \chg{\cite{Hinton2015,Gou2021}}.} \chgold{Distilled runs also use mixup \cite{Zhang2018}. The student and both teachers receive $x_\lambda=\lambda x_a+(1-\lambda)x_b$ with $\lambda\sim\mathrm{Beta}(0.3,0.3)$, and $\mathcal{L}_{\mathrm{CE}}=\lambda\,\mathrm{CE}(\hat p,y_a)+(1-\lambda)\,\mathrm{CE}(\hat p,y_b)$ without label smoothing.} \chgold{Plain runs use label smoothing 0.15 and no mixup, so any distillation effect in Section~\ref{sec:scale} includes these two differences.} \chgold{The larger variants use the settings of Table~\ref{tab:train}.} \chg{Algorithm~\ref{alg:training} summarizes training and evaluation.}

\begin{algorithm}[!t]
\caption{Training and paired evaluation of one model on one test fold}
\label{alg:training}
\small
\begin{algorithmic}[1]
\Require real frames with labels, test fold $f$, model $f_\theta$, optional teachers $T_1,T_2$
\Ensure averaged weights $\bar\theta$, test scores, bootstrap interval against a reference model
\State add augmented copies of every class other than COPD until each class has 9,837 frames
\State draw five class-stratified folds with seed 42, keep the original frames of fold $f$ as test set $\mathcal{T}$, and hold out 10\% of the other frames for validation
\For{epoch $=1,\dots,100$}
  \For{each batch}
    \State apply two time masks of up to 8 columns to all three views
    \State compute $\mathcal{L}$, cross-entropy with $\varepsilon=0.15$, or Eq.~\eqref{eq:kd} with mixup for a distilled student
    \State update $\theta$ with AdamW, clip the gradient norm at 1.0, and update $\bar\theta$ with $\delta_s$
  \EndFor
  \State compute the validation loss of $\bar\theta$
  \If{the validation loss has not fallen by more than $10^{-4}$ in 12 epochs}
    \State stop and keep $\bar\theta$ of the last improvement
  \EndIf
\EndFor
\State score $\bar\theta$ on $\mathcal{T}$ for accuracy, macro F1, MCC, AUC and ECE
\For{$b=1,\dots,B$}
  \State resample the audio files of $\mathcal{T}$ with replacement and re-score both models
\EndFor
\State \Return $\bar\theta$, the scores, and the 2.5th and 97.5th percentiles of the difference
\end{algorithmic}
\end{algorithm}

\begin{table*}[!t]
\centering\small
\caption{\chgold{Training settings. All runs use the balanced pool, two time masks of up to 8 columns, weight averaging and a cosine schedule, and CCT-M and CCT-L train for up to 150 epochs with a 10-epoch linear warm-up, all other models for up to 100.}}
\label{tab:train}
\begin{tabular}{lccccc}
\toprule
Model group & Learning rate & Weight decay & Batch & Label smoothing & Mixup\\
\midrule
\textbf{RespCCT, FreqCCT} & \textbf{0.002946} & \textbf{0} & \textbf{128} & \textbf{0.15} & \textbf{0}\\
Pretrained baselines & $10^{-3}$ (grid for two) & 0 & 128 & 0.15 & 0\\
\chgold{CCT-M plain} & \chgold{0.001414} & \chgold{0.03} & \chgold{256} & \chgold{0.15} & \chgold{0}\\
\chgold{CCT-M with distillation} & \chgold{0.001414} & \chgold{0.03} & \chgold{256} & \chgold{0} & \chgold{0.3}\\
\chgold{CCT-L plain} & \chgold{0.001061} & \chgold{0.05} & \chgold{256} & \chgold{0.15} & \chgold{0}\\
\chgold{CCT-L with distillation} & \chgold{0.001061} & \chgold{0.05} & \chgold{256} & \chgold{0} & \chgold{0.3}\\
\bottomrule
\end{tabular}
\\[2pt]\footnotesize Stochastic depth and MLP dropout are 0.085 and 0.088 for RespCCT and FreqCCT, 0.10 and 0.09 for CCT-M and 0.15 and 0.10 for CCT-L (attention dropout 0.08 and 0.10). Baselines use model defaults.
\end{table*}

\subsection{Five-fold cross-validation and repeated seeds}
\label{sec:cv}
\chgold{One split can flatter a model, so the evaluation used folds and seeds.} \chgold{The frames were divided into five folds stratified by class (seed 42).} \chgold{Fold 0 is the test set for every model comparison (2,359 unmodified frames), and the other four folds supplied training and validation data.} \chgold{Each fold also served in turn as the test set for the two proposed models, and RespCCT was trained on all five folds with three seeds.} Baselines, larger variants and ablations were evaluated on fold 0 with one run each.

\chg{The ablation study removes one component at a time while keeping the recipe of Section~\ref{sec:training}. Ten variants were trained once each on fold 0. Four change the views (mel only, mel with CQT, mel with MFCC, and CQT with MFCC), two change the encoder (no transformer blocks, and attention replaced by a wider feed-forward layer with the same parameter count), two change the pooling (mean over tokens, and a class token), one removes the view gate, and one replaces the tuned recipe with a hand-set reference recipe. Each variant is compared with the full model by the paired bootstrap of Section~\ref{sec:metrics}.}

\subsection{Evaluation metrics and statistical analysis}
\label{sec:metrics}
\chgold{The main metric is the macro-averaged F1 score, which gives each of the $K=7$ classes the same weight whatever its size, a necessary property when 83\% of frames are COPD. With $TP_c$, $FP_c$ and $FN_c$ the true positives, false positives and false negatives of class $c$ (Eq.~\eqref{eq:f1}),}
\begin{equation}
\label{eq:f1}F1_c=\frac{2TP_c}{2TP_c+FP_c+FN_c},\qquad \text{macro F1}=\frac1K\sum_{c=1}^{K}F1_c.
\end{equation}
\chgold{Accuracy, balanced accuracy (mean recall over classes) and the multiclass Matthews correlation coefficient \cite{Gorodkin2004} are also reported.} \chgold{With the confusion matrix $C$, row sums $t_k$, column sums $p_k$, $s=\sum_{ij}C_{ij}$ and $c=\sum_kC_{kk}$ (Eq.~\eqref{eq:mcc}),}
\begin{equation}
\label{eq:mcc}\mathrm{MCC}=\frac{c\,s-\sum_kp_kt_k}{\sqrt{\big(s^2-\sum_kp_k^2\big)\big(s^2-\sum_kt_k^2\big)}}.
\end{equation}
\chgold{The macro one-versus-rest area under the ROC curve (AUC) and the expected calibration error \cite{Guo2017} over ten equal-width confidence bins $S_b$ complete the set (Eq.~\eqref{eq:ece}),}
\begin{equation}
\label{eq:ece}\mathrm{ECE}=\sum_{b=1}^{10}\frac{|S_b|}{n}\,\big|\mathrm{acc}(S_b)-\mathrm{conf}(S_b)\big|.
\end{equation}
\chgold{A file-level vote assigns each audio file the most frequent class among its frames.}

\chgold{Uncertainty in the macro F1 difference between two models on the same test frames is estimated with a paired bootstrap over \chgold{audio files}.} \chgold{\chgold{Audio files} are resampled with replacement, both models are re-scored on each resample, and the 2.5th and 97.5th percentiles of the difference give a 95\% interval.} \chgold{If a rare class is absent from a resample, it is left out of that resample's macro average.} The number of resamples is 500 for the comparisons with pretrained networks, 300 for the ablations and 1,000 for the accuracy differences between larger models. \chgold{A difference is called reliable when its interval excludes zero.} \chgold{These intervals reflect the sampling of test \chgold{audio files} and not the variation between training runs.} \chgold{For accuracy, the McNemar test \cite{McNemar1947} is applied to the discordant pairs, $b$ frames that only model A classifies correctly and $c$ frames that only model B classifies correctly, with statistic $(|b-c|-1)^2/(b+c)$ referred to a chi-square distribution with one degree of freedom.} \chgold{Because McNemar treats frames as independent, the bootstrap is the primary test and McNemar $p$ values are given for reference, adjusted by the Holm step-down procedure when several are reported together.}

\subsection{Explainable AI and model validation}
\label{sec:expl_method}
\chgold{A plausible-looking map can have nothing to do with what the model learned, so the explanations were tested as well as shown.} \chgold{All attribution methods are post hoc and have no effect on training or inference.} \chgold{Two map types were computed for RespCCT on its 16$\times$10 token grid.} \chgold{The contribution map gives each token's share of the class score, its pooling weight times the dot product of its final embedding with the class weight vector.} \chgold{Class-weighted attention rollout multiplies the head-averaged attention matrices of successive blocks with the identity added for the residual connection, $R=\hat A_4\cdots\hat A_1$, and weights the result by the token contributions.}

\chgold{A model randomization check tests whether a map depends on what the model learned.} \chgold{Following Adebayo et al.\ \cite{Adebayo2018}, the trained weights were replaced by random ones layer by layer from the classifier toward the input, and the Spearman rank correlation between the original and the partly randomized map was computed at each step.} \chgold{A map that reflects the trained model should lose this correlation, and a correlation below 0.3 after full randomization was counted as a pass, a threshold chosen by us.} 

\chgold{Two further methods measure which time--frequency regions matter for each class.} \chgold{Occlusion replaces one 4$\times$4 token region in all three views by the per-pixel median of real training frames and records the drop in the predicted class probability.} \chgold{Shapley values \cite{Lundberg2017} were estimated with a permutation explainer over 48 groups (three views by 16 frequency bands), filling masked groups from 64 real training frames.}  \chgold{For each method, the per-class band profiles were correlated pairwise, and the mean off-diagonal correlation measures class specificity, and values near 1 mean that the method gives the same map for every class.}

\chgold{Finally, a logistic regression on the 64-dimensional embeddings before the classifier was trained to predict the database of origin, with five-fold patient-grouped cross-validation, because a model can score well by recognizing the recording source, which no attribution map reveals.}

\subsection{Computational cost assessment}
\chgold{A parameter count does not say what a model costs on a device, so five quantities were measured, namely parameters, floating-point operations per 5 s frame, weight storage in 32-bit floats, peak activation memory and latency.} \chgold{In PyTorch, model latency is the median and 90th percentile of 200 single-frame runs after 30 warm-up runs on one thread of a cloud CPU (Colab).} \chgold{RespCCT was timed without the view gate (59 parameters), and each image transform was timed separately on the same CPU. The total in Table~\ref{tab:cost} adds the median model latency to the 29.6 ms of the three images.} \chgold{It is a sum of separately measured stages, not a measured audio-to-prediction time, and it excludes the enlargement of the images for the pretrained networks.} \chgold{The models were also exported to ONNX and timed with ONNX Runtime 1.31.1 in one Colab session (Intel Xeon 2.20 GHz, one thread, batch size one).} \chgold{The exported graphs reproduced the PyTorch predictions on every test frame.}

\section{Results}
\label{sec:results}

\subsection{Overall classification performance}
\label{sec:overall}
\chgold{All models were scored on the same 2,359 fold-0 test frames, so differences between them come from the models.} \chgold{RespCCT reaches an accuracy of 0.9813 and a macro F1 of 0.9695 (Table~\ref{tab:main}), and a majority vote over the frames of each \chgold{audio file} raises accuracy to 0.985.} \chgold{FreqCCT reaches 0.9758 and 0.9514 with 28\% fewer parameters.} \chgold{Its macro F1 is 0.018 below that of RespCCT (interval $-$0.036 to $-$0.005), so the saving in size has a measurable price.} \chgold{Of the 44 errors of RespCCT, 38 are COPD frames assigned to another class, which gives COPD a recall of 0.981 and a precision of 0.999 (Table~\ref{tab:perclass}).} \chgold{ECE lies between 0.117 and 0.125 for every model. All models were trained with label smoothing of 0.15, which caps the predicted probability near 0.87, so ECE reflects the shared training recipe and is not used to rank the models.} \chg{The macro AUC of RespCCT is 0.9994 (Table~\ref{tab:main}).}

\begin{table*}[!t]
\centering\footnotesize
\setlength{\tabcolsep}{3.5pt}
\caption{\chgold{Fold-0 test results ($n=2{,}359$ frames). $\Delta$F1 is the model's macro F1 minus that of RespCCT with a 95\% paired bootstrap interval, and Vote is \chgold{file-level} majority-vote accuracy. Bold rows are the proposed models.}}
\label{tab:main}
\begin{tabular}{lrrcccccccl}
\toprule
Model & Params & MFLOPs & Acc. & F1 & MCC & Bal. acc. & AUC & ECE & Vote & $\Delta$F1 [95\% CI]\\
\midrule
\textbf{RespCCT} & \textbf{173,187} & \textbf{124.2} & \textbf{.9813} & \textbf{.9695} & \textbf{.9410} & \textbf{.9907} & \textbf{.9994} & \textbf{.117} & \textbf{.985} & \\
\textbf{FreqCCT} & \textbf{124,999} & \textbf{72.5} & \textbf{.9758} & \textbf{.9514} & \textbf{.9244} & \textbf{.9867} & \textbf{.9992} & \textbf{.125} & \textbf{.983} & \textbf{$-$.018 [$-$.036, $-$.005]}\\
ShuffleNetV2 $\times$0.5 & 348,967 & 17.2 & .9784 & .9659 & .9310 & .9842 & .9994 & .121 & .983 & $-$.004 [$-$.014, +.006]\\
MobileNetV3 Small & 931,047 & 24.5 & .9881 & .9777 & .9611 & .9892 & .9983 & .122 & .989 & +.008 [$-$.003, +.020]\\
MobileViT XXS & 953,271 & 109.6 & .9911 & .9843 & .9707 & .9920 & .9999 & .123 & .994 & +.015 [+.006, +.026]\\
\chgold{EfficientNet-B0} & \chgold{4,016,515} & \chgold{168.2} & \chgold{.9915} & \chgold{.9841} & \chgold{.9718} & \chgold{.9901} & \chgold{.9997} & \chgold{.123} & \chgold{.992} & \chgold{+.015 [+.005, +.025]}\\
\chgold{ResNet-18} & \chgold{11,180,103} & \chgold{773.8} & \chgold{.9907} & \chgold{.9808} & \chgold{.9697} & \chgold{.9948} & \chgold{.9998} & \chgold{.121} & \chgold{.993} & \chgold{+.011 [$-$.003, +.023]}\\
\bottomrule
\end{tabular}
\end{table*}

\subsection{Five-fold results and stability}
\chgold{A single split can be favorable, so the result was repeated over folds and seeds.} \chg{With seed 0, RespCCT has a mean accuracy of 0.9760 and a mean macro F1 of 0.9585 across the five folds, with standard deviations of 0.0052 and 0.0106, and FreqCCT has 0.9695 and 0.9408, with standard deviations of 0.0070 and 0.0145. FreqCCT is lower on every fold in both measures, by 0.002 to 0.025 in macro F1 (Table~\ref{tab:folds}).} \chgold{Fold 2 is the hardest for both models (0.9411 and 0.9191), and fold 0 is the easiest for RespCCT, so the single split of Section~\ref{sec:overall} is slightly favorable.} \chg{Because the folds were stratified by class, fold 2 holds the same class mix as the other folds, so its lower score most likely reflects the particular audio files it contains rather than a different class balance.}

\chg{Over three seeds on fold 0, RespCCT reaches an accuracy of 0.9804 $\pm$ 0.0021 and a macro F1 of 0.9653 $\pm$ 0.0037.} \chgold{Over five folds and three seeds (15 runs) it is 0.9594, with an accuracy of 0.9765 and an MCC of 0.9263.} \chgold{Within a fold, the three seeds differ by a mean standard deviation of 0.0056, against 0.0071 between the fold means, so folds and seeds move the score by similar amounts and differences below about 0.01 macro F1 should not be read as real.}  \chg{FreqCCT was repeated over seeds on fold 0 only, with accuracies of 0.9758, 0.9737 and 0.9703 and macro F1 of 0.9514, 0.9416 and 0.9397.}

\begin{table*}[!t]
\centering\small
\caption{\chgold{Five-fold results of RespCCT and FreqCCT (seed 0).}}
\label{tab:folds}
\begin{tabular}{lcccccccc}
\toprule
 & \multicolumn{4}{c}{\textbf{RespCCT}} & \multicolumn{4}{c}{\textbf{FreqCCT}}\\
\cmidrule(lr){2-5}\cmidrule(lr){6-9}
Test fold & Acc. & F1 & MCC & Best epoch & Acc. & F1 & MCC & Best epoch\\
\midrule
0 & .9813 & .9695 & .9410 & 68 of 80 & .9758 & .9514 & .9244 & 47 of 59\\
1 & .9766 & .9625 & .9283 & 61 of 73 & .9675 & .9415 & .8994 & 61 of 73\\
2 & .9674 & .9411 & .8981 & 64 of 76 & .9586 & .9191 & .8757 & 49 of 61\\
3 & .9780 & .9612 & .9299 & 46 of 58 & .9709 & .9358 & .9077 & 56 of 68\\
4 & .9769 & .9584 & .9297 & 84 of 96 & .9749 & .9562 & .9213 & 74 of 86\\
\midrule
Mean & .9760 & .9585 & .9254 & & .9695 & .9408 & .9057 & \\
SD across folds & .0052 & .0106 & .0161 & & .0070 & .0145 & .0196 & \\
\bottomrule
\end{tabular}
\end{table*}

\subsection{Training and validation curves}
\begin{figure*}[!t]
\centering
\includegraphics[width=0.95\textwidth]{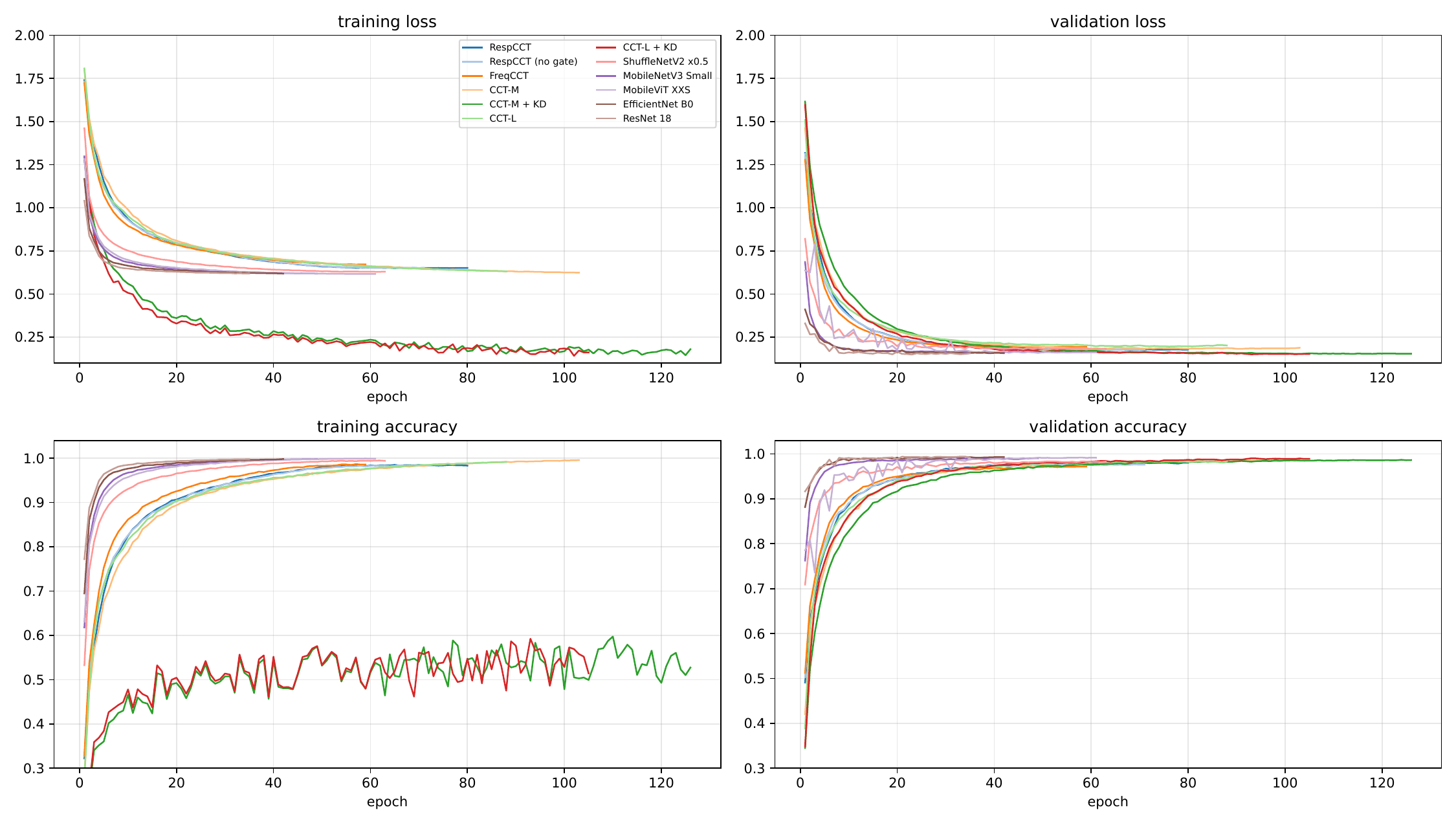}
\caption{\chgold{Training and validation loss (top) and accuracy (bottom) on fold 0 for all twelve runs. For the distilled CCT-M and CCT-L, training accuracy is measured on mixup batches and training loss includes the distillation term, so neither is comparable with the other runs.}}
\label{fig:curves}
\end{figure*}

\chgold{The learning curves show whether each model trained smoothly and when it stopped improving.} \chgold{Figure~\ref{fig:curves} overlays the training and validation loss and accuracy of all twelve fold-0 runs, which are RespCCT with and without its gate, FreqCCT, the five pretrained networks, and CCT-M and CCT-L with and without distillation.} \chgold{Validation loss lies below training loss for every model, as expected when training frames carry label smoothing, time masking, dropout and stochastic depth and validation frames do not.} \chgold{The compact models learn more slowly.} \chgold{RespCCT ran 80 epochs and reached its lowest validation loss of 0.1772 at epoch 68, and FreqCCT ran 59 epochs with its lowest value of 0.1911 at epoch 47.} \chgold{The pretrained networks stopped after 35 to 63 epochs, and ResNet-18 reached 0.1488 at epoch 23 and MobileNetV3 Small 0.1618 at epoch 37.}  \chgold{The curves show that training was stable, but they do not show how a model behaves on recordings outside the pooled data.}

\subsection{Confusion matrices}
\begin{figure*}[!t]
\centering
\includegraphics[width=0.40\textwidth]{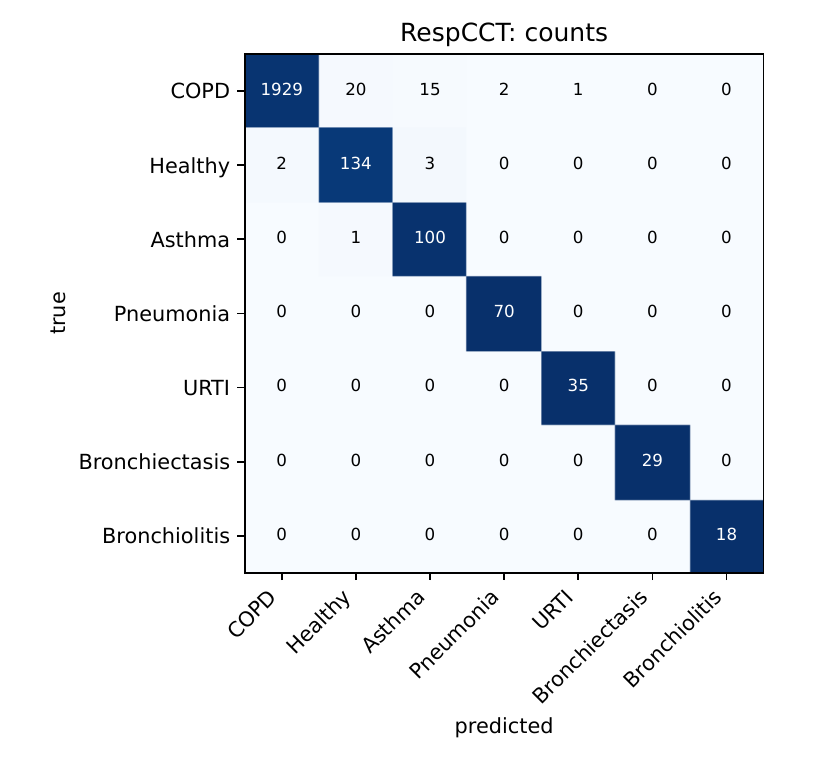}\hfill
\includegraphics[width=0.40\textwidth]{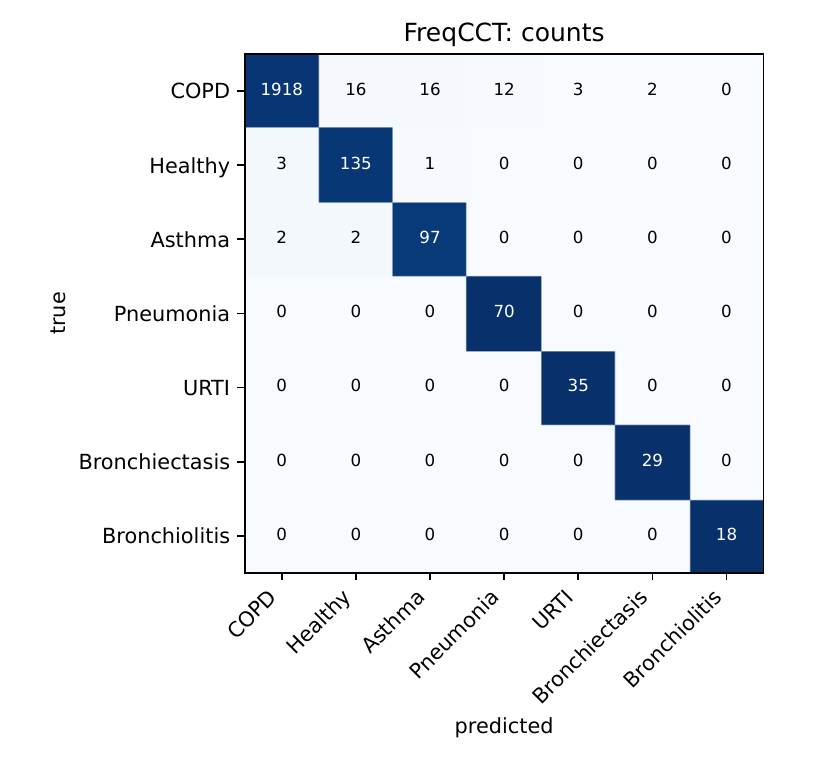}
\caption{\chgold{Confusion matrices (counts) of RespCCT (left) and FreqCCT (right) on the 2,359 fold-0 test frames, with true classes in rows and predicted classes in columns.}}
\label{fig:conf}
\end{figure*}

\chgold{The confusion matrices show where the errors fall (Figure~\ref{fig:conf}).} \chgold{RespCCT makes 44 errors and FreqCCT 57, and in both models most errors are COPD frames assigned to healthy, asthma or pneumonia, because COPD is by far the largest class.} \chg{Of the 38 COPD errors of RespCCT, 20 were assigned to healthy, 15 to asthma, 2 to pneumonia and 1 to URTI, and the 49 COPD errors of FreqCCT follow the same pattern, with 16 going to healthy, 16 to asthma, 12 to pneumonia, 3 to URTI and 2 to bronchiectasis. Healthy and asthma are the two classes drawn mostly from KAUH (588 of 828 and 513 of 520 frames, Table~\ref{tab:data}), so part of these errors may follow the recording source as well as the sound, although the source of each misclassified frame was not analyzed.} \chgold{Both models classify every pneumonia, URTI, bronchiectasis and bronchiolitis frame correctly, so their differences lie in COPD and in the precision losses that its errors cause in the other classes.} 

\subsection{Per-class performance}
\chgold{Macro F1 is sensitive to small classes, where one error changes the score, so the per-class numbers were examined (Table~\ref{tab:perclass}).} \chgold{The 38 COPD errors of RespCCT lower the precision of healthy (0.865) and asthma (0.847), while the recall of every class lies between 0.964 and 1.000.} \chg{In screening terms, the COPD recall of 0.981 means that about 2 of every 100 COPD frames received another label, and the precision of 0.999 means that a frame labeled COPD was almost always COPD.} \chgold{FreqCCT has low precision on pneumonia (0.854) and URTI (0.921), and the 12 and 3 COPD frames it sends to these classes account for most of its lower macro F1.} \chgold{Bronchiectasis and bronchiolitis have no RespCCT errors, but with 29 and 18 test frames from 16 and 13 \chgold{audio files}, perfect scores on these classes carry wide uncertainty.}

\begin{table}[!t]
\centering\small
\setlength{\tabcolsep}{3pt}
\caption{\chgold{Per-class precision (P), recall (R) and F1 on fold 0.}}
\label{tab:perclass}
\begin{tabular}{lrrrrrrr}
\toprule
 & & \multicolumn{3}{c}{\textbf{RespCCT}} & \multicolumn{3}{c}{\textbf{FreqCCT}}\\
\cmidrule(lr){3-5}\cmidrule(lr){6-8}
\chgold{Class} & \chgold{Frames} & \chgold{P} & \chgold{R} & \chgold{F1} & \chgold{P} & \chgold{R} & \chgold{F1}\\
\midrule
\chgold{COPD} & \chgold{1,967} & \chgold{.999} & \chgold{.981} & \chgold{.990} & \chgold{.997} & \chgold{.975} & \chgold{.986}\\
\chgold{Healthy} & \chgold{139} & \chgold{.865} & \chgold{.964} & \chgold{.912} & \chgold{.882} & \chgold{.971} & \chgold{.925}\\
\chgold{Asthma} & \chgold{101} & \chgold{.847} & \chgold{.990} & \chgold{.913} & \chgold{.851} & \chgold{.960} & \chgold{.902}\\
\chgold{Pneumonia} & \chgold{70} & \chgold{.972} & \chgold{1.000} & \chgold{.986} & \chgold{.854} & \chgold{1.000} & \chgold{.921}\\
\chgold{URTI} & \chgold{35} & \chgold{.972} & \chgold{1.000} & \chgold{.986} & \chgold{.921} & \chgold{1.000} & \chgold{.959}\\
\chgold{Bronchiectasis} & \chgold{29} & \chgold{1.000} & \chgold{1.000} & \chgold{1.000} & \chgold{.935} & \chgold{1.000} & \chgold{.967}\\
\chgold{Bronchiolitis} & \chgold{18} & \chgold{1.000} & \chgold{1.000} & \chgold{1.000} & \chgold{1.000} & \chgold{1.000} & \chgold{1.000}\\
\bottomrule
\end{tabular}
\end{table}

\subsection{Comparison with pretrained models}
\label{sec:pretrained}
\chgold{Training from scratch has a cost in accuracy, and Table~\ref{tab:main} shows its size.} \chgold{RespCCT places fifth of seven models, ahead of ShuffleNetV2 $\times$0.5 (0.9659) and FreqCCT (0.9514).} \chgold{MobileViT XXS (0.9843) and EfficientNet-B0 (0.9841) lead it by 0.0148 and 0.0146 macro F1, and both paired intervals exclude zero.} \chgold{MobileNetV3 Small ($+$0.0082) and ResNet-18 ($+$0.0113) are higher on the point estimate, but their intervals include zero, as does that of ShuffleNetV2 (\chgold{$-$0.0036}).} \chgold{RespCCT is therefore reliably below two of the five pretrained networks and not reliably different from the other three.}  

\chgold{The pretrained networks carry weights learned from millions of photographs, while RespCCT and FreqCCT learn from lung sounds alone, and they also store more.} \chgold{RespCCT needs 0.69 MB of weights against 1.40 MB for ShuffleNetV2, 3.72 MB for MobileNetV3 Small, 3.81 MB for MobileViT XXS, 16.07 MB for EfficientNet-B0 and 44.72 MB for ResNet-18, a factor of 2.0 to 64.6 (Table~\ref{tab:cost}), and FreqCCT needs 0.50 MB. No smaller model in the comparison is more accurate than RespCCT, and the next more accurate one, MobileNetV3 Small, has 5.4 times as many parameters.}  

\subsection{Ablation study}
\label{sec:abl}
\begin{figure*}[!t]
\centering
\includegraphics[width=0.92\textwidth]{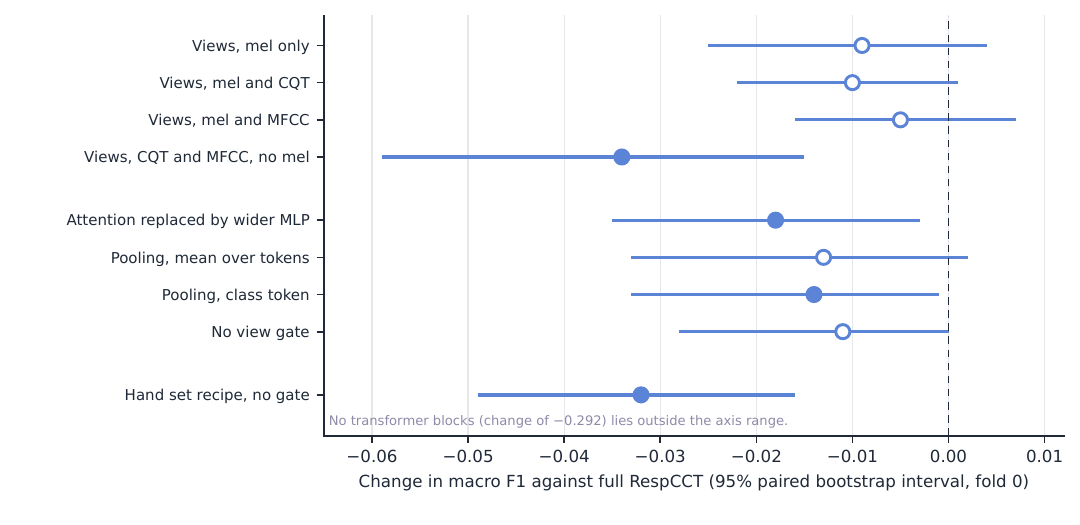}
\caption{\chgold{Change in fold-0 macro F1 of each ablation relative to the full RespCCT, with 95\% paired bootstrap intervals, and filled markers exclude zero. The variant without transformer blocks ($-$0.292) lies outside the axis.}}
\label{fig:abl}
\end{figure*}

\chgold{The ablations show which parts of RespCCT are needed and what can be dropped on a smaller device (Table~\ref{tab:abl}, Figure~\ref{fig:abl}).} \chgold{Each variant removes one component, keeps the recipe of Section~\ref{sec:training} and was trained once, so the intervals describe the sampling of test audio files and not the variation between training runs.}

\chg{Three components carry most of the accuracy of RespCCT. Without the transformer blocks, accuracy falls from 98.13\% to 85.33\% and macro F1 from 0.970 to 0.678, so the tokenizer alone is not enough. Replacing attention by a wider feed-forward layer with the same number of parameters lowers accuracy to 97.41\% and macro F1 by 0.018, with an interval of $-$0.035 to $-$0.003, so attention adds something that capacity alone does not. Removing the mel view, which leaves constant-Q and MFCC, lowers accuracy to 96.65\% and macro F1 by 0.034, with an interval of $-$0.059 to $-$0.015.}

\chg{Most of the remaining changes lower macro F1 by 0.005 to 0.014, and only one of them, the class token, has an interval that excludes zero. Mel alone, or mel with one other view, gives accuracies of 97.75\% to 97.80\% and lowers macro F1 by 0.005 to 0.010, with intervals that include zero, so on this fold the second and third views add a gain that the data cannot confirm. A class token is reliably lower than sequence pooling in macro F1, by 0.014 with an interval of $-$0.033 to $-$0.001 and an accuracy of 97.67\%, and the mean over tokens is lower by 0.013 with an interval that includes zero and an accuracy of 98.01\%.}

\chg{The evidence for the view gate is weaker than for any other component. It adds 0.011 macro F1 in the single ablation run, with an accuracy of 97.92\% without it and an interval that touches zero. Over three seeds on fold 0 the mean macro F1 is 0.9653 with the gate and 0.9613 without it, a difference of 0.004 that lies within the seed spread. The gate was kept because it costs 59 parameters, but the evidence for it is weak. The training recipe has the largest effect after the encoder, since the hand-set reference recipe gives an accuracy of 96.52\% and a macro F1 of 0.938, which is 0.032 below the tuned one, with an interval of $-$0.049 to $-$0.016. Part of the result therefore comes from regularization and weight averaging.}

\begin{table*}[!t]
\centering\small
\caption{\chgold{Ablations of RespCCT on fold 0 (one run each). $\Delta$F1 is the variant minus the full model with a 95\% paired bootstrap interval.}}
\label{tab:abl}
\begin{tabular}{lrcccl}
\toprule
Variant & Params & Acc. & F1 & MCC & $\Delta$F1 [95\% CI]\\
\midrule
\textbf{Full model (sequence pooling, three views, gate)} & \textbf{173,187} & \textbf{.9813} & \textbf{.9695} & \textbf{.941} & \\
Views, mel only & 172,001 & .9780 & .9601 & .929 & $-$.009 [$-$.025, +.004]\\
Views, mel and CQT & 172,594 & .9775 & .9597 & .929 & $-$.010 [$-$.022, +.001]\\
Views, mel and MFCC & 172,594 & .9780 & .9650 & .931 & $-$.005 [$-$.016, +.007]\\
Views, CQT and MFCC (no mel) & 172,594 & .9665 & .9355 & .896 & $-$.034 [$-$.059, $-$.015]\\
Encoder, no transformer blocks & 39,299 & .8533 & .6779 & .581 & $-$.292 [$-$.343, $-$.250]\\
\chgold{Encoder, no attention, wider feed-forward} & \chgold{172,675} & \chgold{.9741} & \chgold{.9516} & \chgold{.919} & \chgold{$-$.018 [$-$.035, $-$.003]}\\
Pooling, mean over tokens & 173,122 & .9801 & .9570 & .937 & $-$.013 [$-$.033, +.002]\\
Pooling, class token & 173,186 & .9767 & .9559 & .927 & $-$.014 [$-$.033, $-$.001]\\
Without view gate & 173,128 & .9792 & .9584 & .934 & $-$.011 [$-$.028, +.000]\\
\chgold{Hand-set reference recipe, no gate} & \chgold{173,128} & \chgold{.9652} & \chgold{.9377} & \chgold{.890} & \chgold{$-$.032 [$-$.049, $-$.016]}\\
\bottomrule
\end{tabular}
\end{table*}

\subsection{Knowledge distillation and ensemble results}
\label{sec:scale}
\chgold{More capacity alone does not close the gap to the pretrained networks (Table~\ref{tab:scale}).} \chg{Without distillation, CCT-M with 0.97 M parameters and CCT-L with 2.15 M reach accuracies of 98.43\% and 97.92\% and macro F1 of 0.968 and 0.964, with 37 and 49 errors, against 98.13\%, 0.970 and 44 errors for the 0.17 M RespCCT. With the two pretrained teachers, MobileViT XXS and MobileNetV3 Small, CCT-M reaches an accuracy of 98.64\% and a macro F1 of 0.975 with 32 errors, and CCT-L 98.52\% and 0.968 with 35 errors.} \chgold{Distilled CCT-M is not reliably different from MobileNetV3 Small (accuracy difference $-$0.0017, interval $-$0.006 to $+$0.0025, $p=0.58$) or from MobileViT XXS ($-$0.0047, interval $-$0.0102 to 0.000, $p=0.082$, Holm-adjusted $p=0.41$)\chgold{,} and its gain over RespCCT ($+$0.0051, interval $-$0.0004 to $+$0.011, $p=0.097$) is not reliable either.} \chgold{Holm adjustment is applied across the six paired comparisons of CCT-M with KD and the ensemble against MobileViT XXS, MobileNetV3 Small and RespCCT.}

\chgold{The ensemble of the four larger CCTs makes 23 errors (accuracy 0.9903, macro F1 0.9813).} \chgold{It cannot be separated from MobileViT XXS ($p=0.84$) or MobileNetV3 Small ($p=0.40$), and it is reliably better than RespCCT ($+$0.0089, interval $+$0.0035 to $+$0.0142, Holm-adjusted $p=0.008$), at 6.2 M parameters and 3.3 GFLOPs.}  \chgold{Because the distilled runs also use mixup and no label smoothing, the 0.007 macro F1 gain of CCT-M cannot be credited to the teachers alone, and the larger models train 2.7 to 5.3 times longer than RespCCT (54 to 106 min against 20).} \chgold{Reaching the level of the pretrained networks with this architecture therefore costs 5.6 to 36 times the parameters of RespCCT.}

\begin{table*}[!t]
\centering\small
\caption{\chgold{Larger CCTs, distillation and ensembling on fold 0. KD means distilled from MobileViT XXS and MobileNetV3 Small with mixup and no label smoothing, and the ensemble averages CCT-M and CCT-L with and without KD.}}
\label{tab:scale}
\begin{tabular}{lrrccccc}
\toprule
Model & Params & MFLOPs & Acc. & F1 & MCC & Errors & Vote\\
\midrule
\textbf{RespCCT} & \textbf{173,187} & \textbf{124.2} & \textbf{.9813} & \textbf{.9695} & \textbf{.941} & \textbf{44} & \textbf{.985}\\
\chgold{CCT-M} & \chgold{967,619} & \chgold{536.8} & \chgold{.9843} & \chgold{.9681} & \chgold{.950} & \chgold{37} & \chgold{.987}\\
\chgold{CCT-M + KD} & \chgold{967,619} & \chgold{536.8} & \chgold{.9864} & \chgold{.9748} & \chgold{.956} & \chgold{32} & \chgold{.988}\\
\chgold{CCT-L} & \chgold{2,151,811} & \chgold{1,135.5} & \chgold{.9792} & \chgold{.9637} & \chgold{.933} & \chgold{49} & \chgold{.982}\\
\chgold{CCT-L + KD} & \chgold{2,151,811} & \chgold{1,135.5} & \chgold{.9852} & \chgold{.9680} & \chgold{.952} & \chgold{35} & \chgold{.987}\\
Ensemble of four CCTs & 6.2 M & 3,345 & .9903 & .9813 & .968 & 23 & .991\\
MobileNetV3 Small & 931,047 & 24.5 & .9881 & .9777 & .961 & 28 & .989\\
MobileViT XXS & 953,271 & 109.6 & .9911 & .9843 & .971 & 21 & .994\\
\bottomrule
\end{tabular}

\end{table*}

\subsection{Explainable AI results}
\label{sec:expl}
\chgold{The explanation tests ask whether the maps depend on the trained model, whether they differ between classes, and what the learned features encode besides the diagnosis.} \chgold{The first two tests used 242 test frames, up to 40 per class, on which RespCCT reaches an accuracy of 0.988.}

\chg{A map that depends on the trained model should change when the weights are replaced by random ones. When the trained weights were replaced layer by layer from the classifier toward the input, the class-weighted rollout kept a rank correlation of 0.95 with the original map after the classifier was reset and 0.66 after block 2, and the correlation fell to 0.10 after block 1 and ended at 0.17, below the threshold of 0.3, so the rollout passes the check. The contribution map ended at 0.012, but it had already fallen to 0.216 when only the classifier was reset, which suggests that it reflects the classification head more than the encoder.}

\begin{figure*}[!t]
\centering
\includegraphics[width=0.8\textwidth]{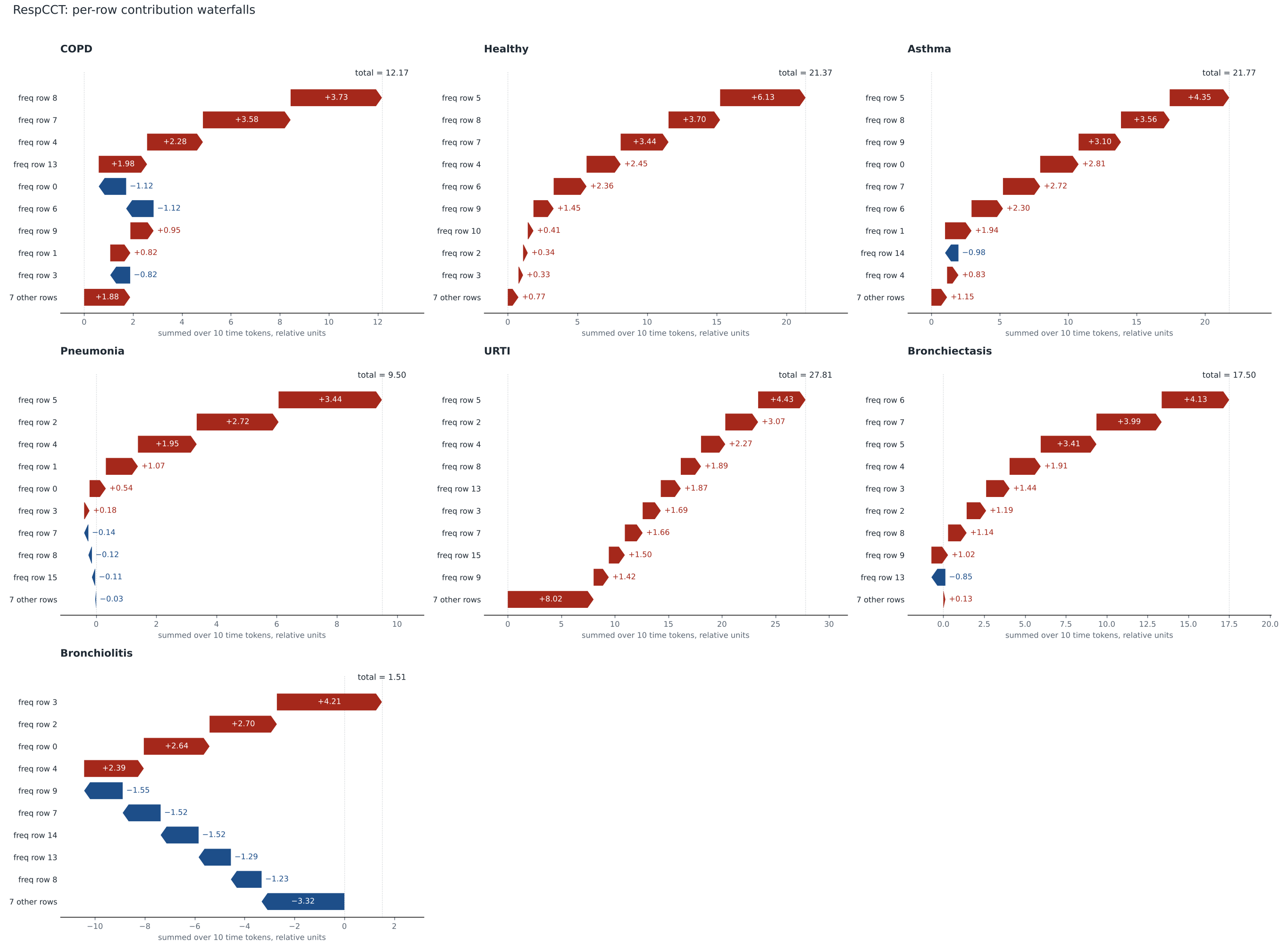}
\caption{\chgold{Per-row contribution waterfalls of RespCCT for each class, built from the occlusion maps summed over time tokens. Red rows lower the class probability when occluded and blue rows raise it. Row 5 dominates Healthy and Pneumonia, Asthma and URTI spread over several rows, and Bronchiolitis ends near zero because its red and blue rows cancel.}}
\label{fig:occl}
\end{figure*}

\begin{figure*}[!t]
\centering
\includegraphics[width=0.75\textwidth]{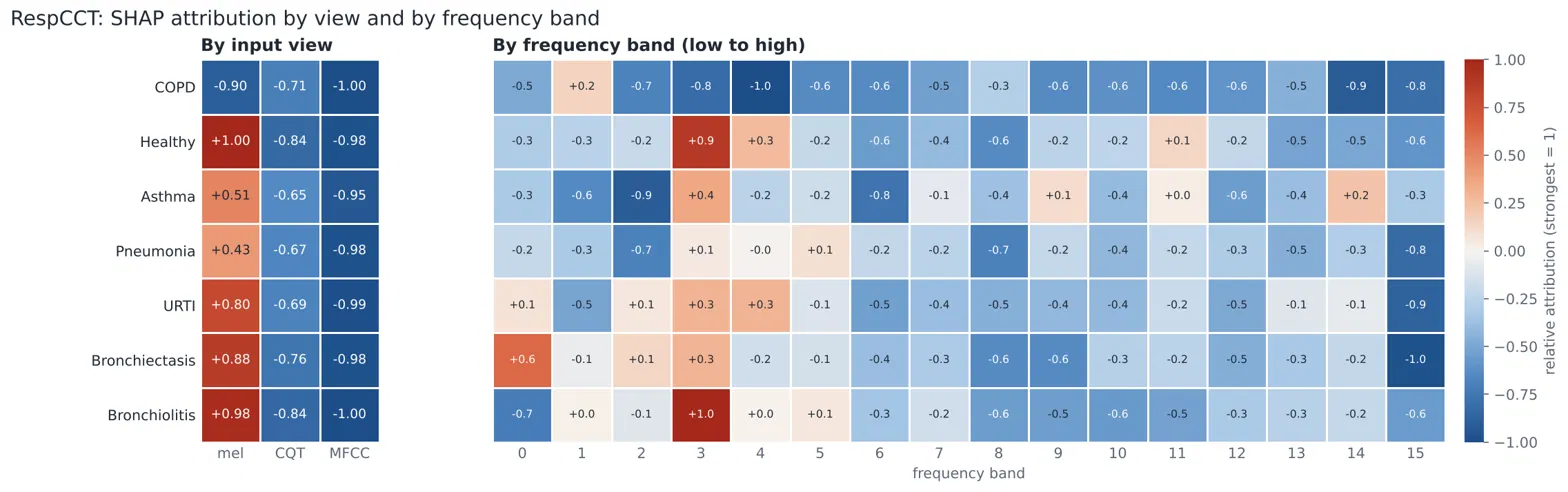}
\caption{\chgold{Shapley attribution of RespCCT per class, grouped by view (left) and by frequency band from low to high (right).}}
\label{fig:shap}
\end{figure*}

\chg{Occlusion and Shapley values give different maps for different classes, as Figures~\ref{fig:occl} and \ref{fig:shap} show. The mean correlation between the band profiles of two classes is 0.292 for occlusion and 0.277 for Shapley values in RespCCT, and 0.223 and 0.390 in FreqCCT. A value near 1 would mean that every class receives the same map. In both models the healthy and pneumonia maps concentrate on row 5, while the asthma, URTI and COPD maps spread over several bands.}

\chg{The three input views do not contribute equally, but the two measures used here disagree on how unequal they are. The learned gate ranks the views in the same order for every class, with mean weights of about 0.99 for mel, 0.45 for CQT and 0.27 for MFCC, and the per-frame Shapley values of Figure~\ref{fig:shap} give the same order, with mel highest in six of seven classes and MFCC lowest in all seven. A Shapley decomposition of the macro F1, which measures what each view adds to the final score, gives shares of 0.49 for mel, 0.47 for CQT and 0.04 for MFCC. Both measures agree that MFCC adds little and disagree on CQT, so neither settles how much each view contributes. The first MFCC coefficient carries the energy of the frame and dominates the min--max scaling of its image, so the other 29 rows span a narrow range, which may explain the low weight of MFCC.}

\chg{Converted to hertz, the COPD attribution places its largest share in the lowest mel band, 26 to 103 Hz with 28.5\%, followed by 129 to 206 Hz with 14.7\%. No clinical meaning is attached to these bands, because the main component of heart sounds lies between 20 and 150 Hz \cite{Molaie2015}, and the stethoscope or the recording room may contribute as much as the disease.}

\chg{RespCCT changes few predictions when the input is shifted in time. Circular shifts of 0.6, 1.3 and 2.5 s changed 1.3, 1.8 and 1.7\% of the predictions, so its evidence lies in which local patterns occur and not in where they sit within the frame.}

\label{sec:probe}\chg{The recording source can be read from the learned features. A classifier on the 64-dimensional embedding of RespCCT predicts the database of origin with an accuracy of 0.958 under patient-grouped cross-validation, and 0.964 for FreqCCT, against 0.582 for always naming the largest source. Because the sources differ in class mix, as Table~\ref{tab:data} shows, part of this information concerns the label itself, and whether the predictions change once source cues are removed was not tested.}

\subsection{Computational cost, model size and inference time}
\begin{figure*}[!t]
\centering
\includegraphics[width=\textwidth]{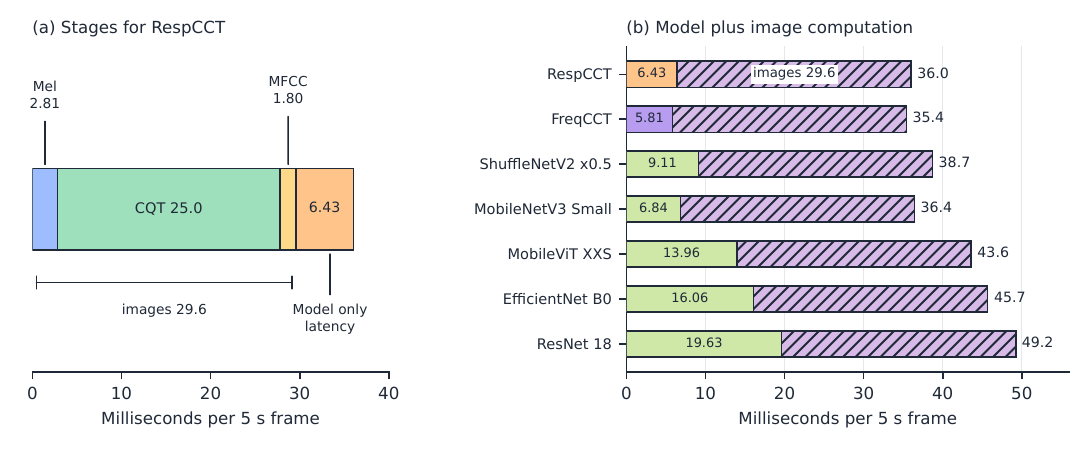}
\caption{\chgold{Cost per 5 s frame on one CPU thread, with (a) the three image transforms and the RespCCT model and (b) model and image time (hatched) for each network.}}
\label{fig:cost}
\end{figure*}

\chgold{Cost decides whether a model runs on a device at all, so it is reported next to accuracy (Tables~\ref{tab:cost} and \ref{tab:onnx}, Figure~\ref{fig:cost}).} \chgold{Arithmetic and memory point in different directions.} RespCCT needs 124.2 MFLOPs per frame, 5.1 times MobileNetV3 Small (24.5) and 7.2 times ShuffleNetV2 (17.2), and its peak activation memory (655 kB) is 3.6 times that of MobileNetV3 Small (184 kB). \chgold{In PyTorch on one CPU thread, RespCCT takes 6.43 ms per frame and FreqCCT 5.81 ms, against 6.84 ms for MobileNetV3 Small and 9.11 to 19.63 ms for the other four pretrained networks.} \chgold{The margin over MobileNetV3 Small comes from one machine and shrinks to about 1\% once the 29.6 ms of image computation is added (36.0 against 36.4 ms).} \chgold{The pretrained networks also read inputs with four times as many pixels, the form for which they were designed, so this is a comparison of deployable systems and not of architectures at equal input size.}

\begin{table*}[!t]
\centering\small
\setlength{\tabcolsep}{4pt}
\caption{\chgold{Cost per 5 s frame in PyTorch on one Colab CPU thread. Latency is the median / 90th percentile, and Total adds the 29.6 ms of the three images (sum of separately measured stages). RespCCT is timed without its view gate.}}
\label{tab:cost}
\begin{tabular}{lrrrrcr}
\toprule
Model & Params & MFLOPs & Weights (kB) & Peak act. (kB) & Model latency (ms) & Total (ms)\\
\midrule
\textbf{RespCCT (without gate)} & \textbf{173,128} & \textbf{124.2} & \textbf{692.5} & \textbf{655.4} & \textbf{6.43 / 7.21} & \textbf{36.0}\\
\textbf{FreqCCT} & \textbf{124,999} & \textbf{72.5} & \textbf{500.0} & \textbf{655.4} & \textbf{5.81 / 6.57} & \textbf{35.4}\\
ShuffleNetV2 $\times$0.5 & 348,967 & 17.2 & 1,395.9 & 245.8 & 9.11 / 11.57 & 38.7\\
MobileNetV3 Small & 931,047 & 24.5 & 3,724.2 & 184.3 & 6.84 / 7.71 & 36.4\\
MobileViT XXS & 953,271 & 109.6 & 3,813.1 & 327.7 & 13.96 / 16.14 & 43.6\\
\chgold{EfficientNet-B0} & \chgold{4,016,515} & \chgold{168.2} & \chgold{16,066.1} & \chgold{983.0} & \chgold{16.06 / 17.95} & \chgold{45.7}\\
\chgold{ResNet-18} & \chgold{11,180,103} & \chgold{773.8} & \chgold{44,720.4} & \chgold{655.4} & \chgold{19.63 / 21.93} & \chgold{49.2}\\
\bottomrule
\end{tabular}
\end{table*}

\begin{figure}[!t]
\centering
\includegraphics[width=\columnwidth]{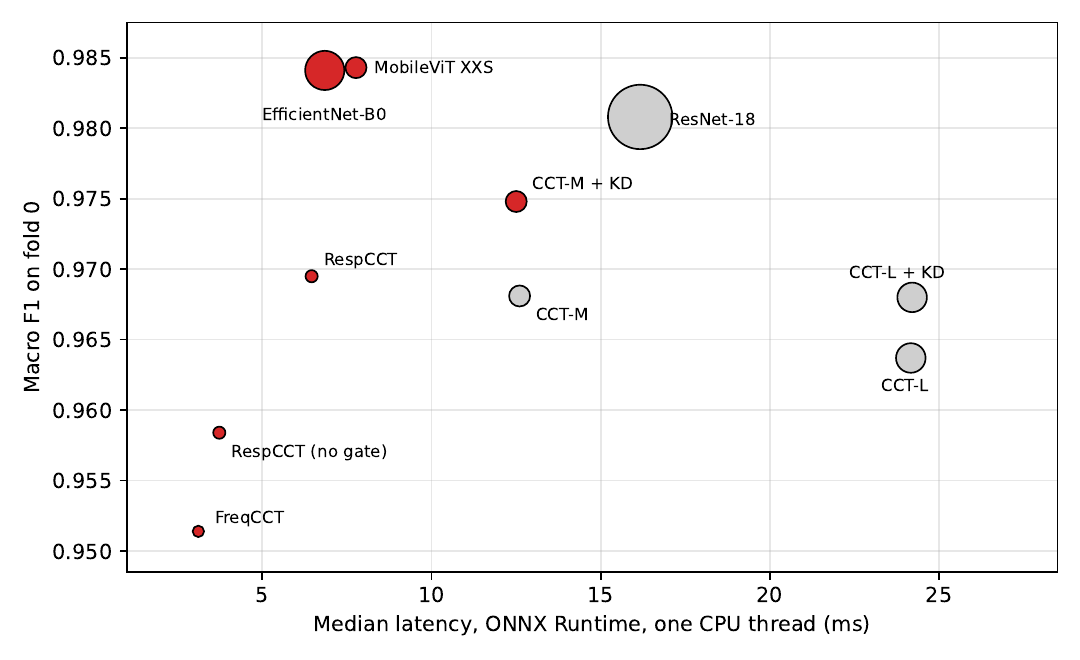}
\caption{\chg{ONNX Runtime latency on one CPU thread against fold-0 macro F1 for the ten models of Table~\ref{tab:onnx}, with RespCCT shown with and without its gate. Bubble area shows file size, and red marks models that no faster and smaller model exceeds in macro F1.}}
\label{fig:pareto}
\end{figure}

\chgold{The ONNX Runtime session gives a second, independent timing (Table~\ref{tab:onnx}).} \chgold{FreqCCT ran at 3.11 ms per frame and RespCCT (timed without the gate) at 3.73 ms, against 6.85 ms for EfficientNet-B0, 7.77 ms for MobileViT XXS and 16.17 ms for ResNet-18.} \chgold{These three networks therefore need 2.2 to 5.2 times the time of FreqCCT. The exported files are 0.56 MB and 0.74 MB, against 3.84 to 42.64 MB for these three networks.} \chg{The same graph with its view gate measured 6.46 ms (99th percentile 49.5 ms) against 3.73 ms (6.29 ms) without it, and both gave the same predictions. The gap was not investigated, which is why the latency of RespCCT is reported for the version without the gate. Figure~\ref{fig:pareto} plots macro F1 against latency and file size for the ten models of the session.}

\begin{table}[!t]
\centering\small
\setlength{\tabcolsep}{3pt}
\caption{\chgold{ONNX Runtime 1.31.1 on one CPU thread (Intel Xeon 2.20 GHz, one Colab session, batch size one), showing file size, median latency and latency relative to FreqCCT.}}
\label{tab:onnx}
\begin{tabular}{lrrr}
\toprule
\chgold{Model} & \chgold{File (MB)} & \chgold{Median (ms)} & \chgold{Ratio}\\
\midrule
\chgold{\textbf{FreqCCT}} & \chgold{\textbf{0.559}} & \chgold{\textbf{3.11}} & \chgold{\textbf{1.00}}\\
\chgold{\textbf{RespCCT (without gate)}} & \chgold{\textbf{0.742}} & \chgold{\textbf{3.73}} & \chgold{\textbf{1.20}}\\
\chgold{\textbf{RespCCT (with gate)}} & \chgold{\textbf{0.743}} & \chgold{\textbf{6.46}} & \chgold{\textbf{2.08}}\\
\chgold{EfficientNet-B0} & \chgold{15.314} & \chgold{6.85} & \chgold{2.20}\\
\chgold{MobileViT XXS} & \chgold{3.840} & \chgold{7.77} & \chgold{2.50}\\
\chgold{CCT-M} & \chgold{3.813} & \chgold{12.60} & \chgold{4.05}\\
\chgold{CCT-M + KD} & \chgold{3.813} & \chgold{12.50} & \chgold{4.02}\\
\chgold{ResNet-18} & \chgold{42.640} & \chgold{16.17} & \chgold{5.20}\\
\chgold{CCT-L} & \chgold{8.330} & \chgold{24.17} & \chgold{7.77}\\
\chgold{CCT-L + KD} & \chgold{8.330} & \chgold{24.20} & \chgold{7.78}\\
\bottomrule
\end{tabular}
\end{table}

The three images take longer to compute than the network takes to read them. \chgold{On the Colab CPU the mel spectrogram costs 2.81 ms, the CQT 25.0 ms (84\% of the total) and the MFCC 1.80 ms, 29.6 ms in all (Figure~\ref{fig:cost}a), against 6.43 ms for the RespCCT model.} The separately measured stages sum to about 36 ms per frame, and a new frame starts every 2.5 s, a duty cycle of 1.4\% if the stages add as measured. \chgold{Dropping the CQT would save 25.0 ms at a macro F1 change of $-$0.005 (mel with MFCC, interval $-$0.016 to $+$0.007, Table~\ref{tab:abl}), a result that has not been checked on other folds.}

\begin{table*}[!t]
\centering\footnotesize
\setlength{\tabcolsep}{3pt}
\caption{\chg{Results reported by earlier studies on diagnostic classification of lung sounds and by the present models. The data, classes and protocols differ, so the values are not ranked, and n/a means that the value is not given in the source text checked.}}
\label{tab:prior}
\begin{tabular}{>{\raggedright\arraybackslash}p{2.3cm}>{\raggedright\arraybackslash}p{2.7cm}>{\raggedright\arraybackslash}p{2.3cm}>{\raggedright\arraybackslash}p{3.2cm}>{\raggedright\arraybackslash}p{1.5cm}>{\raggedright\arraybackslash}p{3.0cm}}
\toprule
Study & Data and classes & Evaluation & Reported result & Parameters & Timing\\
\midrule
Pham et al. \cite{Pham2021} & ICBHI, 3 disease classes & official split and five-fold cross-validation & ICBHI Score 84.0 (official split), 90.5 (five folds) & n/a & n/a\\
Zhang et al. \cite{Zhang2024} & ICBHI, 5,054 augmented samples & random split & accuracy 99.01\% & n/a & n/a\\
Hassan et al. \cite{Hassan2024} & ICBHI and KAUH, 3 s frames & as reported & accuracy 0.997 (ICBHI), 1.0 (KAUH), 0.998 (combined) & two convolution layers & n/a\\
Roy et al. \cite{Roy2024} & ICBHI, KAUH and RD@TR, 7 classes & five-fold cross-validation & accuracy 98.88\%, sensitivity 98.27\% & 0.414 M & 3.48 s per segment on a Raspberry Pi 4\\
Park et al. \cite{Park2025} & ICBHI and KAUH & 80/20 split & accuracy 79.68\% (GPU), 89.46\% (Hailo-8 NPU) & 264 K & 245.9 ms on the NPU\\
Dong et al. \cite{Dong2026} & ICBHI, and KAUH with RD@TR, COPD against other & as reported & accuracy 99.23\% (ICBHI), 98.48\% (KAUH with RD@TR) & n/a & n/a\\
RespCCT (this study) & ICBHI, KAUH and RD@TR, 7 classes & five folds and three seeds & accuracy 97.65\%, macro F1 0.959 & 0.173 M & 36.0 ms per frame on one CPU thread, sum of stages\\
FreqCCT (this study) & ICBHI, KAUH and RD@TR, 7 classes & five folds, seed 0 & accuracy 96.95\%, macro F1 0.941 & 0.125 M & 35.4 ms per frame, same set-up\\
\bottomrule
\end{tabular}
\end{table*}

\section{Discussion}
\label{sec:disc}

\chgold{The central question was how far a small transformer can go when it learns from lung sounds alone, and what that costs in memory, arithmetic and time.} \chgold{The answer is that it comes close but does not reach the best pretrained networks.} \chgold{MobileViT XXS and EfficientNet-B0 lead RespCCT by 0.015 macro F1 with intervals that exclude zero, and they reach accuracies of 99.11\% and 99.15\% against 98.13\%.} \chg{An ensemble of four larger CCTs reaches an accuracy of 99.03\% and a macro F1 of 0.981 at 6.2 M parameters.} \chgold{RespCCT, however, cannot be separated from MobileNetV3 Small, ResNet-18 and ShuffleNetV2 on 2,359 frames.} \chgold{It needs no ImageNet weights and stores 0.69 MB where the pretrained networks store 1.4 to 44.7 MB. The gap of 0.015 therefore comes with the absence of pretrained weights and a storage reduction of 2.0 to 64.6 times.}

\chgold{The two models suit different devices.} \chgold{RespCCT is the better choice where a phone, a gateway or a board can hold 173,187 parameters (0.69 MB) and run 124.2 MFLOPs per frame, and its macro F1 is higher than that of FreqCCT (0.9695 against 0.9514, interval of the difference excluding zero).} \chgold{FreqCCT suits devices where memory is the main limit.} \chgold{It has 124,999 parameters (0.50 MB) and needs 42\% fewer operations, and in both timing set-ups its model latency was below that of every pretrained network timed alongside it (5.81 ms in PyTorch, 3.11 ms in ONNX Runtime), at a cost of 0.018 macro F1.}  \chgold{These device labels are a reading of parameter counts, operations and two CPU timings, not tests on the devices themselves.}

\chg{\textbf{Fewer errors or a smaller model.} Fewer errors matter most in a medical setting, and the two strongest pretrained networks make 21 and 20 errors on the test fold against 44 for RespCCT and 57 for FreqCCT, as Table~\ref{tab:scale} shows. The type of error matters as well, because a disease frame labeled healthy is the costliest mistake in screening. RespCCT makes 21 of them among 2,220 non-healthy frames, which is 0.95\%, FreqCCT 18 or 0.81\%, MobileViT XXS 11 or 0.50\% and EfficientNet-B0 6 or 0.27\%, while ShuffleNetV2, the pretrained network closest in size, makes 24 or 1.08\%. The intervals of these rates for RespCCT and MobileViT XXS overlap, at 0.62 to 1.44\% and 0.28 to 0.89\%, but they treat frames as independent, so a paired file-level test is needed before the difference is called real.}

\chg{On accuracy alone the pretrained networks are ahead, and this study does not claim otherwise. The case for the compact models rests on what they need to run. RespCCT stores 0.69 MB of weights against 3.81 MB for MobileViT XXS and 16.07 MB for EfficientNet-B0, its network takes 3.73 ms in ONNX Runtime without the view gate against 7.77 and 6.85 ms, and it does not depend on ImageNet weights. These figures hold at 32-bit precision on a shared cloud CPU. Whether they survive integer quantization of every model, a paired test of the missed-disease rate and timing on the target device remains to be shown.}

\chgold{The multi-view result is weaker than expected.} \chgold{Removing mel is the only change with a reliable cost, and mel with MFCC loses 0.005 with an interval that includes zero.} \chgold{Because the CQT takes 84\% of the image computation, mel with MFCC is a candidate configuration, but it rests on one fold.}  

\chg{The architecture ablations show which parts carry the accuracy that remains.} \chgold{A feed-forward layer with the same parameter count is 0.018 lower, so attention adds something that capacity alone does not.} \chgold{RespCCT has no positional embedding and changes few predictions under circular time shifts, which indicates that its evidence lies in which local patterns occur rather than in where they sit within the 5 s frame.} \chgold{Sounds tied to a strict breathing phase may behave differently and were not tested.} 

 \chgold{Distillation brought CCT-M level with MobileNetV3 Small, but together with mixup and without label smoothing, so the teachers cannot be credited alone.}

\chg{Table~\ref{tab:prior} places the present models beside six earlier studies on diagnostic classification of lung sounds. Their reported accuracies lie between 79.68\% and 100\%, but the databases, number of classes, splits and metrics differ, so the values are not ranked. The closest setting is that of Roy et al.\ \cite{Roy2024}, who used the same three databases and seven classes. Two of their results differ from ours. Their fine-tuned pretrained networks scored below their compact network, whereas here the pretrained networks lead RespCCT, and their networks with one or two views fell below the three-view network, whereas in the view ablations of Section~\ref{sec:abl} only the removal of mel cost a reliable amount of macro F1. One possible explanation is that their network processes each view in a separate stream, whereas RespCCT stacks the views as channels of one tokenizer, but this explanation was not tested.}

\chgold{Not every attribution map passed the checks.} \chgold{Class-weighted rollout and contribution maps lose their dependence on the trained weights under full randomization, but the contribution map follows the classifier more than the encoder, and rollout separates the classes less than occlusion and Shapley values do.} \chgold{The embedding also identifies the source database with 0.958 accuracy, so part of what the network learns is where a frame was recorded.} \chgold{Because the sources differ in class mix, this information can raise pooled accuracy without helping in a new clinic.} \chgold{The predictions are therefore decision support for \chgold{audio files} similar to those of the three databases and never a diagnosis, and the attribution tests show dependence on the trained model, not clinical meaning.}

\section{Limitations and future work}
\label{sec:limits}

\chg{The study has four main limitations. First, the folds were drawn at the frame level and consecutive frames overlap by 2.5 s, so a test frame can share audio with a training frame of the same audio file, and the scores describe new frames from the pooled cohort and not new patients. Second, the source database can be read from the learned features with an accuracy of 0.958 while the sources differ in class mix, so part of the pooled accuracy may come from source cues. Third, the baselines, larger variants and ablations were run once on fold 0 and distillation was combined with mixup, so the gap of 0.015 macro F1 and the distillation result rest on single runs. Fourth, no model was timed on a phone or a single-board computer, the timings come from shared cloud CPUs, and the ONNX graph of RespCCT with its view gate ran at 6.46 ms against 3.73 ms without it for a reason that was not investigated.}

\chg{Each limitation leads to one next step. A file-grouped evaluation, a patient-grouped evaluation and a test on a held-out database would show how far the macro F1 falls once no audio is shared with training. A source-balanced test set, a leave-one-database-out evaluation and adversarial removal of source information, as in the gradient reversal of Kim et al.\ \cite{Kim2024}, would test whether the probe accuracy can approach the 0.582 of the majority source while the macro F1 is kept. Repeating the comparison over seeds and folds, distilling with and without mixup, and adding a respiratory foundation model such as OPERA \cite{Zhang2024OPERA} as a further reference would show whether the gap of 0.015 to MobileViT XXS and EfficientNet-B0 keeps its interval away from zero. Timing both models with integer quantization on a phone and a low-cost board, tracing the cost of the view gate, and testing mel with MFCC as a cheaper input on all folds, since removing the constant-Q transform would save 25.0 ms of the 29.6 ms spent on the images, would give the measured time per frame and the change in accuracy and macro F1.}

\section{Conclusion}
\label{sec:concl}

\chg{This study asked how far a compact transformer can go when it learns from lung sounds alone and what it costs to run. On 2,359 test frames of seven classes, RespCCT with 173,187 parameters and 0.69 MB of weights reached an accuracy of 98.13\% and a macro F1 of 0.970, and 97.65\% and 0.959 over five folds and three seeds, while FreqCCT with 124,999 parameters and 0.50 MB reached 97.58\% and 0.951. MobileViT XXS and EfficientNet-B0 lead RespCCT with accuracies of 99.11\% and 99.15\% and a macro F1 of 0.984, and the intervals of these gaps in macro F1 exclude zero, whereas the other three pretrained networks cannot be separated from RespCCT. RespCCT stores 2.0 to 64.6 times less weight than the five pretrained networks. Distillation from two pretrained teachers brought CCT-M to an accuracy of 98.64\%, but its gain over RespCCT is not reliable and cannot yet be separated from the change in training recipe. The largest cost lies in front of the network, because the three input images need 29.6 ms per frame against 6.43 ms for the RespCCT network, so a device implementation should start with the images. Class-weighted rollout passed the randomization check and occlusion and Shapley maps differ between classes, yet the learned features identify the source database with an accuracy of 0.958, so part of the pooled accuracy may come from a source cue. RespCCT suits devices that can hold about 0.7 MB and FreqCCT suits devices with tighter memory, in both cases as decision support for audio files similar to those of the three databases. While the results are promising, further evaluation across unseen recording environments and independent databases would help establish broader generalization. Testing on actual edge devices would also provide a clearer picture of practical efficiency.
Future work will focus on held out database testing, model quantization, and reducing input processing time for efficient deployment on low cost devices.}

\section*{Ethics statement}
\chgold{This study used only public, de-identified \chgold{audio files}, and no new patient data were collected, so no ethical approval was required.} \chgold{The models are a research aid for recorded lung sounds and not a diagnostic device. Any clinical use would need prospective validation, review of every prediction by a clinician, and approval under the medical-device regulations of the country concerned.}

\section*{CRediT authorship contribution statement}
\textbf{Auntor Chakma:} Methodology, Software, Writing (original draft), Supervision. \textbf{Mst. Farhana Akter:} Code and analysis, fine-tuning, literature review. \textbf{Abu Henaf Rashid Ahmad Mufti:} Visualization, interpretation, writing and manuscript editing. \textbf{Asmana Yasmin:} Proof reading, dataset overview and manuscript editing.

\section*{Declaration of generative AI and AI-assisted technologies in the manuscript preparation process}
\chgold{During the preparation of this work the authors used Paperpal and Grammarly to edit the language and finalize the structure of the text.} \chgold{After using these tools, the authors reviewed and edited the content as needed and take full responsibility for the content of the published article.}

\section*{Declaration of competing interest}
The authors declare that they have no known competing financial interests or personal relationships that could have appeared to influence the work reported in this paper.

\section*{Funding}
\chgold{This research did not receive any specific grant from funding agencies in the public, commercial, or not-for-profit sectors.}

\section*{Data availability}
The ICBHI, KAUH and RD@TR databases are public \cite{Rocha2019,Fraiwan2021,Altan2020} and are available from \cite{ICBHIData2017}, \cite{KAUHData2021} and \cite{RDTRData2020}, respectively. The code, configuration files, per-run results and figure sources of this study are available at https://github.com/auntor101/respcct-freqcct under the MIT license. Trained model weights are not released.

\bibliographystyle{elsarticle-num}
\bibliography{refs}

\end{document}